\documentclass[fleqn,usenatbib]{mnras}
\usepackage{newtxtext,newtxmath}

\usepackage{amssymb}
\usepackage{graphicx}
\usepackage{amsmath}
\usepackage[T1]{fontenc}
\usepackage{ae,aecompl}
\usepackage{tabularx}
\usepackage{multirow}

\title[A wide-area search for the first galaxies]
{A search for the first galaxies across ${\mathbf{>0.6\,deg^2}}$ of JWST imaging: new evidence for a rapid decline in star-formation activity at ${\mathbf{z>12}}$}
\author[D. J. ~McLeod et al.]
{D. J. McLeod$^{1}$\thanks{Email: derek.mcleod@ed.ac.uk}, J. S. Dunlop$^{1}$, R. J. McLure$^{1}$,  C. T. Donnan$^{2}$, R. Begley$^{10}$, S. Antonogiannaki$^{1}$, \newauthor D. Magee$^{3}$, G. D. Illingworth$^{3}$, P. Arrabal Haro$^{4,5}$, C. Bondestam$^{1}$, A. C. Carnall$^{1}$, F. Cullen$^{1}$,\newauthor M. Dickinson$^{2}$, R. S. Ellis$^{6}$, B. L.  Frye$^{7}$, H. Golawska$^{1}$, N. A. Grogin$^{8}$, I. J. B. Holst$^{1}$, P. S. Kamieneski$^{9}$, \newauthor H.-H. Leung$^{1}$, F.-Y. Liu$^{1}$, T. M. Stanton$^{1}$, E. R. Tittley$^{1}$
\footnotesize\\
$^{1}$Institute for Astronomy, University of Edinburgh, Royal Observatory, Edinburgh EH9 3HJ\\
$^{2}$ NSF's National Optical-Infrared Astronomy Research Laboratory, 950 N. Cherry Avenue, Tucson, AZ 85719, USA\\
$^{3}$ Department of Astronomy and Astrophysics, UCO/Lick Observatory, University of California, Santa Cruz, CA 95064, USA\\
$^{4}$ Center for Space Sciences and Technology, UMBC, 5523 Research Park Dr, Baltimore, MD 21228 USA\\
$^{5}$ Astrophysics Science Division, NASA Goddard Space Flight Center, 8800 Greenbelt Rd, Greenbelt, MD 20771, USA\\
$^{6}$ Department of Physics and Astronomy, University College London, Gower Street, London WC1E 6BT, UK\\
$^{7}$ Department of Astronomy/Steward Observatory, University of Arizona, 933 N. Cherry Avenue, Tucson, AZ 85721, USA\\
$^{8}$ Space Telescope Science Institute, 3700 San Martin Drive, Baltimore, MD 21218, USA\\
$^{9}$ Department of Physics \& Astronomy, Chalmers University of Technology, SE-412 96 Gothenburg, Sweden\\
$^{10}$ Armagh Observatory and Planetarium, College Hill, Armagh, BT61 9DG, N. Ireland, UK}
\date{Accepted XXX. Received YYY; in original form ZZZ}
\pubyear{2026}

\begin{document}
\label{firstpage}
\pagerange{\pageref{firstpage}--\pageref{lastpage}}
\maketitle

\begin{abstract}
We present a new determination of the evolving galaxy UV luminosity function (LF) over the extreme redshift range $12.5~<~z~<~18.5$, based on a wide-area search of $>$\,$0.6$\,deg$^2$ of \textit{JWST} NIRCam imaging containing $>$\,$150$ independent sight-lines. We find evidence for an accelerated decline in the UV LF, and hence inferred star-formation rate density ($\rho_{\rm SFR}$), over the $\simeq100\,\rm{Myr}$ cosmic time interval between $z=11$ and $z=13.5$. Moreover, based on a notable lack of galaxy candidates at $z>14.5$, we find evidence for an even more rapid descent in star-formation activity towards earlier times, with our new measurement of $\rho_{\rm SFR}$ at $z\simeq15.5$ lying significantly below an extrapolation of the log-linear $\rho_{\rm SFR}(\rm z)$ relation inferred from early \textit{JWST} LF studies. Instead, we find that the evolution in $\rho_{\rm SFR}(\rm z)$ at these very early times is better described by a piece-wise log-linear relation, in which the decline in $\rho_{\rm SFR} (\rm z)$ at $z>12$ is $\simeq4$ times steeper than at redshifts $z < 12$. Our observational results are consistent with a number of theoretical models of galaxy evolution which have incorporated a range of treatments in an attempt to explain the prevalence of UV-bright galaxies at least out to $z \simeq 12$ (e.g., increased star-formation efficiency, stochastic star-formation histories, an evolving stellar initial mass function and/or a shift towards attenuation-free stellar populations). However, our results are also entirely consistent with a relatively simple galaxy evolution model with no such adjustments, in which the rapid evolution of the dark-matter halo mass function at early times is for a while partially masked by progressively younger stellar ages, with the inferred epoch of first galaxy formation lying at  $z\simeq15$.

\end{abstract}

\section{INTRODUCTION}
\label{section1}
In only its first four years of science operations, the {\it James Webb Space Telescope} ({\it JWST}) has already transformed and clarified our picture of early galaxy evolution. In particular, a clear consensus has rapidly emerged that the galaxy UV luminosity function (LF), and hence cosmic star-formation rate density ($\rho_{\mathrm{SFR}}$), declines only gradually from redshift $z \simeq 8$ (the effective limit of {\it Hubble Space Telescope} ({\it HST}) studies) back to $z \simeq 11-12$, less than 400\,Myr after the Big Bang  (e.g., \citealt{Donnan2023a,Donnan2023b,donnan2024,mcleod2024,Perez-Gonzalez2023,Finkelstein2023,finkelstein2024}; but see \citealt{willott2024}). Inevitably, such statistical population studies rely extensively on galaxy {\it photometric} redshifts, but these results are now supported by {\it spectroscopic} confirmation (with NIRSpec on {\it JWST}) of dozens of $z>10$ galaxies \citep{Harikane2023b,robertsborsani2024}, including some particularly UV-bright ($M_{\rm UV}\leq-20$) examples \citep{castellano2024,napolitano2025,kokorev2025b,Donnan2025b}. There are now spectroscopic confirmations as distant as $z \simeq 14$ \citep{carniani2024}, with the current record holder at $z_{\rm spec}=14.44$ \citep{naidu2025}.

Although anticipated by some pre-{\it JWST} observational studies (e.g., \citealt{McLeod2016}) the swift
discovery of an abundant population of ultra high-redshift galaxies with \textit{JWST} came as a surprise to many (e.g., \citealt{Oesch2018}) and undoubtedly runs counter to the expected rapid evolution of the dark matter halo mass function at such early times. Unsurprisingly, therefore, these new results have triggered a flurry of activity focused 
both on potential biases in the data obtained so far, as well as plausible physical explanations of the observed early emergence of the UV-bright galaxy population. 

In terms of potential bias, with the initial {\it JWST} discoveries of galaxies at $z \ge 10$ being derived from the relatively small survey areas (and hence modest cosmological volumes) covered by the Early Release Science (ERS) NIRCam programmes, the potential issue of field-to-field variance has been raised and discussed 
by a number of authors (e.g., \citealt{Adams2024, willott2024}), a concern highlighted by the remarkable number of bright $z>10$ galaxies discovered in the Abell2744 region alone (\citealt{Castellano2023, napolitano2025}). 

However, the relatively gradual high-redshift evolution of the galaxy UV LF inferred from the early, small-area {\it JWST} CEERS and GLASS imaging (e.g., \citealt{bouwens2023,Donnan2023a,Finkelstein2023,finkelstein2024,Adams2024}) has now been confirmed by studies based on wider+deeper high-quality NIRCam imaging obtained in {\it JWST} Cycles 1 and 2. In particular, \citet{donnan2024} analysed a combined $\simeq400$ arcmin$^2$ of 8-band NIRCam imaging spanning $\simeq 2$\,dex in depth, by utilising the PRIMER (\citealt{dunlop2021}; Dunlop et al. in prep), JADES \citep{eisenstein2023} and NGDEEP \citep{Bagley2023b} surveys (along with an improved statistical methodology) to produce the most robust measurement of the early evolution of the UV LF to date, broadly confirming (but also refining) the early {\it JWST} findings. Reassuringly, very similar results have also been obtained  by \citet{mcleod2024} from a compilation of many independent NIRCam pointings covering a total effective area of $\simeq210$\,arcmin$^2$. Most recently, \citet{Weibel2025} combined $\simeq0.28$\,deg$^2$ of imaging from several extragalactic legacy fields plus pure-parallel NIRCam observations from PANORAMIC \citep{williams2025}, finding similar trends at least out to $z\simeq10-11$.

The persistence of a relatively high, integrated UV luminosity density ($\rho_{\rm UV}$) from the galaxy population out to extreme redshifts is accompanied by (and in part driven by) an apparent change in the shape of the UV LF at $z > 6-7$, in the sense that the bright-end appears to evolve particularly slowly, and hence the high-redshift UV LF is better described by a double power-law than a Schechter function (which has a steeper exponential decline at the bright end) during the first Gyr. This 
trend was previously hinted at in pre-\textit{JWST} wide-area ground-based studies (e.g., \citealt{Bowler2020,Kauffmann2022}) and in large compilations of legacy and pure-parallel \textit{HST} imaging (e.g., \citealt{Leethochawalit2023,FinkelsteinCANDELS,BagleyHST}), with proposed explanations including less efficient mass quenching and/or declining dust attenuation at early times. This change in the shape of the UV LF helps to explain the ease with which early, relatively shallow {\it JWST} NIRCam studies were able to uncover significant numbers of $z \simeq 10$ galaxies. 

The observed slow decline of the galaxy UV LF at $z > 8$, and in particular the persistence of relatively bright galaxies back to early times, have prompted a number of authors to propose various modifications to pre-existing theoretical predictions of early galaxy evolution. For example, the inclusion of increased UV variability from bursty star formation (e.g., \citealt{Mason2023,shen2023}), the adoption of higher star-formation efficiencies (e.g., \citealt{Harikane2023,Harikane2025}) as perhaps produced by feedback-free starbursts \citep{Dekel2023,Li2024}, and/or the assumption  of a top-heavy stellar initial mass function (IMF) at the highest redshifts (e.g., \citealt{Hutter2025}) can all help to explain, at least in part, the new extreme-redshift UV LF results. However, the physical foundations for (and cosmological timing of) some of these proposed modifications remain weak, and while a sudden transition to more bursty star-formation histories in the first Gyr can help to explain the change in the shape of the UV LF, it cannot reproduce the persistently high integrated UV luminosity density, $\rho_{\rm UV}$ (and hence the inferred high-redshift behaviour of $\rho_{\rm SFR}$). 

As an alternative to the proposed changes in star-formation physics in the first Gyr, others have argued that the observed evolution of the UV LF (and $\rho_{\rm UV}$) at extreme redshift
can in fact be explained naturally by stellar populations of progressively younger average age \citep{Donnan2025a,schaye2025}. This equates to an assumption of an increasing UV-light:mass ratio as one approaches the putative epoch of first galaxy/star formation
(which in the model of \citet{Donnan2025a} has to lie at $z \simeq 15$) rather than increased star-formation efficiency. Ultimately, this degeneracy can only be broken (and the debate settled) by measurements of the galaxy stellar mass function (GSMF) out to extreme redshifts, but certainly out to $z \simeq 8$ there is as yet no direct evidence for increased star-formation efficiency. Finally, the attenuation-free model introduced by \citet{Ferrara2023}, in which dust is believed to be driven away from so-called ``blue monster'' galaxies in outflows (see also \citealt{ziparo2023, ferrara2025}), has also been shown to reproduce the observed UV LF at $z>10$ and, indeed, studies of galaxy UV continuum slopes by \citet{cullen2023,cullen2024} provide evidence for attenuation-free stellar populations at $z>11$, although there have been recent examples of apparently dusty $z_{\rm spec}\geq10$ sources being discovered \citep{Donnan2025b,rodighiero2026}.

While debate continues over the correct explanation of the slow evolution of the UV-bright galaxy population back to $z \simeq 12$, the observational redshift frontier has rapidly moved on to $z\gtrsim13$. Here, despite the remarkable progress afforded by \textit{JWST} thus far, we are still firmly in the regime of small number statistics, with only modest numbers of photometrically-selected candidates \citep{bouwensUDF2023,austin2023,robertson2024,whitler2025} and even fewer spectroscopic confirmations \citep{Curtis-Lake2023,wang2023,naidu2025}. 

Nonetheless, by pushing the deepest NIRCam imaging available to its detection limits, some authors have reported a handful of extreme-redshift galaxy candidates beyond $z \simeq 15$. For example, faint F200W dropout candidates have been identified in the GLIMPSE imaging over the Abell S1063 field  \citep{kokorev2025}, while \citet{perezgonzalez2025} have claimed the discovery of several F277W dropouts in the combined NGDEEP+MIDIS+MIDIS-RED imaging survey. Such galaxies, if confirmed to lie at the claimed redshifts $z\simeq16-25$, would indicate significant star-formation activity as early as $\simeq$100-200\,Myr after the Big Bang. However, at present, none of these candidates can be regarded as robust (due to the lack of dynamic range in the imaging), with the application of any realistic cosmological prior favouring alternative lower-redshift solutions. 

In contrast to such reports of extreme-redshift objects, the pure-parallel BEACON survey has yielded zero $z\geq13$ galaxy candidates across $\simeq$180\,arcmin$^2$ of NIRCam imaging \citep{morishita2025}, while the \citet{donnan2024} LF, based on one of the largest \textit{JWST}-based searches to date, predicts a cumulative $p(z)$ at $z \simeq 14.5$ equivalent to only $\simeq$1.3 galaxies. Moreover, the aforementioned \citet{Weibel2025} study finds evidence for a rapid decline in $\rho_{\rm UV}$ beyond $z\simeq12$. Given the relative ease with which {\it JWST} has discovered galaxies out to $z \simeq 12$, it is tempting to interpret this apparent deficit of $z\gtrsim13$ galaxies as a sign that the major epoch of first galaxy formation lies around $z\simeq 15$, as inferred independently from the "constant efficiency" model of \citet{Donnan2025a}.

It is important to note that this new debate over the galaxy number density at $z \gtrsim 13$ is fundamentally different 
from the earlier struggle (roughly a decade ago) to properly establish the evolution of the galaxy UV LF at $z > 8$ with {\it HST} \citep{mcleod2015,McLeod2016,Ishigaki2018,Oesch2018}.
This is because the 1.6$\,\mu$m long-wavelength limit of WFC3 on {\it HST} effectively precluded the robust selection of galaxies at $z \gtrsim 9$, barring a few notable exceptions \citep{Bouwens2011,Coe2013,oesch2014,Oesch2016}. By contrast, the search for dropout galaxy candidates at $z \simeq 13-15$ is not compromised by the extensive red wavelength coverage provided by NIRCam; in principle, with {\it JWST}, galaxies should be nearly as easy to discover at $z \simeq 15$ as at $z \simeq 13$. 

Nonetheless, concerns remain that the failure to securely identify galaxies at $z \simeq 15$ with {\it JWST} could, at least in part, be a consequence of insufficient depth, area and/or wavelength resolution across the Cycle 1-2 NIRCam surveys. The width and location of the NIRCam long-wavelength (LW) channel broadband filters can create additional challenges in selecting robust high-redshift candidates without the assistance of additional deep medium-band imaging, especially when trying to avoid contamination by extreme emission-line galaxies at more modest redshifts, as with the pathological case of CEERS-93316 \citep{Donnan2023a,ArrabalHaro2023}. Deeper imaging in the short-wavelength (SW) NIRCam bands such as F090W, F115W and F150W is also desirable for identifying more secure Lyman breaks at $\lambda_{\rm rest}=1216\,$\AA\,(see the discussion by \citealt{adams2025}). Finally, wide-area datasets spanning a large number of non-contiguous fields are important for mitigating concerns over field-to-field variance, and thus pure-parallel data sets such as PANORAMIC \citep{williams2025} and BEACON \citep{morishita2025} offer a potentially lucrative avenue for constructing samples of ultra high-redshift galaxy candidates. 

In an attempt to resolve the remaining uncertainty over very early galaxy evolution, this study seeks to establish whether there exists a significant population of galaxies at $z \gtrsim 12.5$ by leveraging a uniquely large dataset spanning $>$150 sight-lines that together cover $>0.6$\,deg$^2$ on the sky. This large combined survey search area allows us to place statistically significant constraints on the number density of ultra high-redshift galaxies across a wide dynamic range in UV luminosity, largely free from the concerns over field-to-field variance uncertainties which hampered the early \textit{JWST} searches. To help with the search for fainter galaxies, as well as including ultra-deep ``blank'' fields (e.g., JADES, NGDEEP), we extend our analysis to deep fields that benefit from gravitational lensing by massive foreground clusters (e.g., GLIMPSE, CANUCS, PEARLS, VENUS). By searching for NIRCam F150W dropouts across this unique dataset, this study aims to provide a complete census of robust $z>12.5$ candidates, and to constrain the evolving UV LF (and hence $\rho_{\rm SFR}$) over the redshift range $12.5<z<18.5$, reaching back to within $\simeq$\,200\,Myr of the Big Bang.

This paper is structured as follows. In Section \ref{section2}, we provide a description of the assembled multi-field NIRCam dataset, and then in Section \ref{section3} we describe the process of catalogue construction and subsequent high-redshift galaxy candidate selection. In Section \ref{section4}, we present and discuss our final ultra high-redshift galaxy sample. Section \ref{section5} describes the determination of the galaxy UV luminosity function over the extreme redshift range $12.5<z<18.5$, and the resulting inferred early evolution of star-formation rate density is in Section \ref{section6}. Our conclusions are summarised in Section \ref{section7}. Throughout this paper, all magnitudes are given in the AB system \citep{Oke1974,Oke1983}, and we adopt a flat cosmology with $\Omega_{0}=0.3, \Omega_{\Lambda}=0.7$ and $H_{0}=70$ kms$^{-1}$Mpc$^{-1}$.

\section{DATA}
\label{section2}
Here we provide an overview of the combined multi-field dataset that was employed in this study. Specifically, we compiled a wide range of publicly available high-quality \textit{JWST} NIRCam imaging data from the Early Release Observations (ERO), the Early Release Science (ERS) programmes, and a number of Cycle 1-4 GTO and GO surveys that can, in principle, be leveraged to select galaxies at $z>12$. The key criteria for a dataset to be included in this analysis were imaging in F150W (the key drop-out filter for galaxy selection in the redshift range of interest), plus NIRCam imaging in at least three filters longward of this bandpass (including one of F182M, F200W or F210M). These criteria enable the robust identification of the Lyman break at $z>12$ along with reasonable constraints on the UV continuum slope, $\beta$ (where $f_{\lambda} \propto \lambda^{\beta}$).

We note that for a subset of cluster fields included in this search, e.g. from the recent public VENUS survey (PID 6882; PIs Fujimoto, Coe), there are not yet publicly available lensing maps. While we report any robust $z>12$ candidates that we uncover from these fields, these candidates (and the corresponding survey area in which they are found) are not included in the LF calculations. This is because we do not have adequate constraints on the effective survey area of these fields, nor do we have magnification values to reliably determine the intrinsic UV luminosities ($M_{1500}$) of these particular galaxies. We indicate which lensed fields this applies to in Tables \ref{table: depths}-\ref{table: depths3}.

The sheer number of sight-lines included in our combined survey allows us to essentially eliminate field-to-field variance uncertainty (also known as `cosmic variance') in our UV LF analysis, as this scales down with $\sqrt{N}$, where N is the number of sightlines \citep{Trenti2008}. It is only at the very faint end of the UV LF, where the number of fields able to probe to very low luminosities dramatically decreases, that this could begin to contribute significantly to our uncertainty budget (albeit these lower luminosity galaxies are expected to be less strongly clustered). However, it is important to note that some of the $\>150$ sightlines are contiguous, as would be expected when NIRCam parallel pointings are produced while following up previous NIRCam imaging with other {\it JWST} instruments such as NIRSpec. And, while the NGDEEP and JADES GOODS-South deep imaging surveys represent different sightlines, they are undoubtedly going to be correlated in their galaxy demographics. Moreover, by their design, the pure-parallel surveys have some pointings that inevitably overlap with legacy fields such as PRIMER, CEERS and JADES, due to the fact these early {\it JWST} imaging survey fields contain many sought-after targets for follow-up observations with NIRSpec (see, e.g., ``CEERS-East'').

Unless specified otherwise in the detailed description below, we utilise our own new reduction of each NIRCam imaging dataset using the PRIMER Enhanced NIRCam Image-processing Library reduction package (\textsc{PENCIL}; Magee et al. in prep), which is a customised version of the standard \textit{JWST} pipeline. There are inevitably some small variations in the pipeline version (and the associated CRDS context) depending on when each dataset became available and was reduced. However, the vast majority ($\gtrsim90\%$) of the NIRCam imaging datasets were reduced with pipeline version 1.13.4 or greater (about 10\% are pipeline version 1.10.2), including the latest wisp templates (version 3; August 2024) from STScI. The rest are public data releases produced by the respective survey teams as indicated (CEERS, PRIMER, JADES-South). The typical pmap used is $\geq1300$ across the $\simeq$150 datasets reduced with \textsc{PENCIL}. For a small subset of the data ($<10\%$), it was found that in the short-wavelength bands, even with the latest wisp templates, there remained some residual scattered light. This was further improved in these images using an additional \textsc{Source-Extractor} background subtraction.

\subsection{Updates to previously analysed NIRCam datasets}
As part of the recent studies by \citet{mcleod2024} and \citet{cullen2024}, we amassed a sample of $z>9.5$ galaxy candidates from a raw survey area of $\simeq320$\,arcmin$^2$, spread across fifteen \textit{JWST} data sets. These are re-analysed here, with some significant differences, which we briefly summarise below.

For the Abell 2744 field, where we previously combined the data from DDT 2756 (PI Chen) and UNCOVER (PI Labbe; \citealt{Bezanson2022}), we have now added in the NIRCam imaging from more recent \textit{JWST} programmes targeting this cluster: All the Little Things (ID 3516; PI Matthee), MAGNIF (PID 2883; PI Sun), PID-3538 (PI Iani) and Medium Bands, Mega Science (PID-4111; PI Suess) provide a wealth of additional filter coverage in F070W, F090W and numerous medium bands, enabling excellent spectral energy distribution (SED) sampling. We use the UNCOVER public data release \citet{Furtak2023} lens model.

Two other cluster fields that were previously analysed, MACS 0647 (ID 1433; PI Coe) and WHL 0137 (ID 2282; PI Coe), each now benefit from an additional epoch of imaging in one of the NIRCam modules. The RXJ 2129 cluster previously imaged in DDT2767 (PI Kelly) has now also been targeted by the recent VENUS imaging, which we include. The GLASS survey (ID 1324; PI Treu; \citealt{treu2022}) recently received an additional epoch of observations, further increasing the depth in two NIRCam modules. The ERO field, SMACS 0723 (\citealt{Pontopiddan2022}) has also now been augmented, in this case by new F090W, F115W and F444W imaging from PID-4043 (PI Witten). For SMACS 0723, we use the \citet{Pascale2022} lens model.

The \textit{JWST} Medium Deep Fields NEP TDF (ID 2738; PI Windhorst; \citealt{Windhorst2023}) now includes several extra epochs of data, providing more survey area to the west and southwest. We use the CEERS team's public data release for the full CEERS footprint \citep{Bagley2023}, with the associated ancillary \textit{HST} F435W, F606W and F814W data \citep{Grogin2011,Koekemoer2011,wang2025}. To the east of CEERS (``CEERS-East''), there is a fortuitous combination of pure-parallel survey pointings (Sapphires, BEACON), and NIRCam parallel imaging taken during NIRSpec observations of CEERS (CAPERS PID 6368, PID 4287), which we also now include (using our own reduction).

As part of \cite{cullen2024}, the imaging from NGDEEP (ID 2079; \citealt{Bagley2023b}) and JADES DR1 (ID 1180, 1210; \citealt{rieke2023,eisenstein2023}) was analysed. For the former, we now include both epochs of NGDEEP observations, as well as MIDIS and MIDIS-RED (ID 1283; 6511 PI Oestlin), which combine to provide exceptional depth and partial medium-band coverage. In GOODS-South, we have also updated the imaging to reflect the second data release from JADES \citep{eisenstein2023,eisenstein2025}, which includes an additional $\simeq40$\,arcmin$^2$. Ancillary \textit{HST} data in GOODS-South comes from the Hubble Legacy Fields \citep{Illingworth2016} release. We had also previously analysed the JEMS (PID-1963; PI Williams) portion of JADES GOODS-South \citep{Donnan2023b}, where two $z>12$ candidates were uncovered (see also \citealt{bouwensUDF2023}). Finally, our search also now includes additional area from programmes ID 1286 (PI Luetzgendorf) and DDT6541 (PI Egami).

Although they have not recently received any additional imaging, we also perform a re-analysis of the remaining four fields previously included in \cite{mcleod2024}: Cartwheel, Stephan's Quintet (PI Pontoppidan; \citealt{Pontopiddan2022}) and J1235 (ID 1063; PI Sunnquist). These benefit from improved reductions using the updated version of the \textsc{PENCIL} pipeline (with the associated updated CRDS context).

\subsection{Summary of additional NIRCam datasets}
We add a wealth of other new extragalactic \textit{JWST} NIRCam surveys that have now become publicly available, including the other GTO JADES field, GOODS-North (PID 1181; \citealt{eisenstein2023}), for which we utilise our own independent reduction.

The PRIMER survey (\citealt{dunlop2021}; Dunlop et al. in prep) covers the remaining two {\it HST} CANDELS \citep{Grogin2011,Koekemoer2011} fields, COSMOS and UDS, providing deep imaging over $\simeq$390\,arcmin$^2$. We include the full-area, internal v1 data release imaging over both fields (Magee et al. in prep).

We include the CANUCS (PI Willott; e.g. \citealt{willott2024}) imaging over five cluster fields, as well as their associated parallel fields. While individually of modest area, these fields feature a useful combination of depth and medium-band coverage. In particular, three of these fields (MACS 0416, MACS 1149 and Abell 370), are Hubble Frontier Fields \citep{Lotz2017}, which have also been imaged in numerous other programmes (including, e.g, Technicolour; PID 3362: PI Muzzin) which augment the depth, area and spectral coverage further. We include the Technicolour broad bands and medium bands for the two fields taken earlier, MACS 0416 and Abell 370. The lensing models for these fields were those released in CANUCS DR1 \citep{sarrouh2026}.

The GTO PEARLS programme (PID 1176; PI Windhorst) provides imaging of numerous cluster fields in eight NIRCam bands, including the aforementioned MACS 0416 and MACS 1149. One PEARLS target, PLCK G165.7+67.0 (hereafter PLCK G165; \citealt{frye2024}), has had several additional epochs of imaging taken as part of other programmes: PID 4446 (PI Frye) and PID 4744 (PIs Frye, Pierel).

As part of GLIMPSE \citep{atek2025}, Abell S1063 was observed with ultradeep NIRCam imaging with exceptional sensitivity across nine NIRCam bands from F090W to F480M, and has previously been found to contain $z>15$ candidates (\citealt{kokorev2025}, see Section \ref{section: comparison_candidates}). We include our reduction of this data, and use the public Hubble Frontier Fields lensing model \citep{zitrin2009,zitrin2013}. The last Frontier Field to be imaged by \textit{JWST} was MACS 0717, as part of VENUS. We use the public CATS lens model (\citealt{jauzac2012,richard2014}; see also \citealt{Jullo2009}).

As well as MACS 0717, VENUS includes many new cluster fields not previously imaged with \textit{JWST}, such as MACS 0257-2325 (e.g., \citealt{nakane2025}). For the clusters we analysed that were previously \textit{HST} CLASH clusters \citep{Postman2012}, we used the public lens models as described in \citet{zitrin2009,zitrin2013,Zitrin2015}. For the subset of clusters that were previously \textit{HST} RELICS \citep{Coe2019} clusters, we used their publicly released lens models (see \citealt{Oguri2010,cerny2018,paterno-mahler2018,acebron2018,acebron2019,acebron2020,cibirka2018}). For the SGAS clusters (e.g. from TEMPLATES: \citealt{rigby2025}, and PID 4125; PI Florian) we used the lens models described in \citet{sharon2020,sharon2022,florian2021}. Finally, the lens models for MACS 2135 and PEARLS CLG 1212 were taken from \citet{zitrin2016} and \citet{zitrin2020}, respectively.

A large component of our search area comes from wide-area pure-parallel imaging from the Cycle 1 survey PANORAMIC (PID 2514; \citealt{williams2025}) and the Cycle 2 survey BEACON (PID 3990; \citealt{morishita2025}). These combine to give a similar overall area to the legacy fields PRIMER+JADES+CEERS, but with the added advantage of providing mostly non-contiguous pointings on the sky, alleviating the effects of field-to-field variance. A potential limitation with these surveys is that the depths in the drop-out filters are typically significantly shallower than those redward of the Lyman break. However, this is mitigated to a large extent by our requirement of 8$\sigma$ detections in the longer-wavelength filters (see Section \ref{section3}). 

We exclude from our search those BEACON pointings that lack F200W coverage, or any F182M or F210M imaging to fill the gap in spectral coverage at $\simeq2\,\mu$m. In Cycle 3, two additional pure-parallel programmes were undertaken: Sapphires (PID 6434; PI Eigami) and POPPIES (PID 5398; PI Kartaltepe). Although these were predominantly grism surveys, some pointings included NIRCam imaging observations that we have included in our search. Finally, we include a small amount of COSMOS-Web \citep{Casey2023} imaging that overlaps with these pure-parallel pointings to provide more spectral coverage and sensitivity.

\begin{figure}
    \centering
    \includegraphics[width=0.5\textwidth]{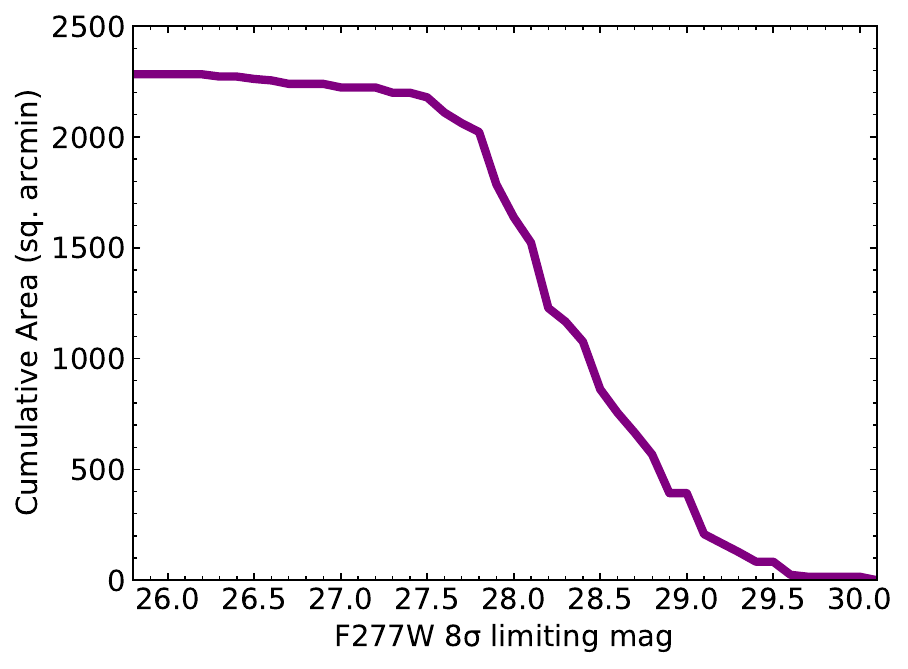}
    \caption{The cumulative raw survey area searched as a function of the global 8$\sigma$ limiting magnitude in F277W. The limiting magnitude is calculated in a 0.20$^{\prime\prime}-$diameter aperture and corrected to total assuming a point-source correction. Note that $>70\%$ of the area is as deep as 28.0 mag, providing a powerful combination of wide-area and depth with which to search for the first galaxies.}
    \label{fig:Areas_Depths}
\end{figure}

\begin{figure*}
    \centering
    \includegraphics[width=0.8\textwidth]{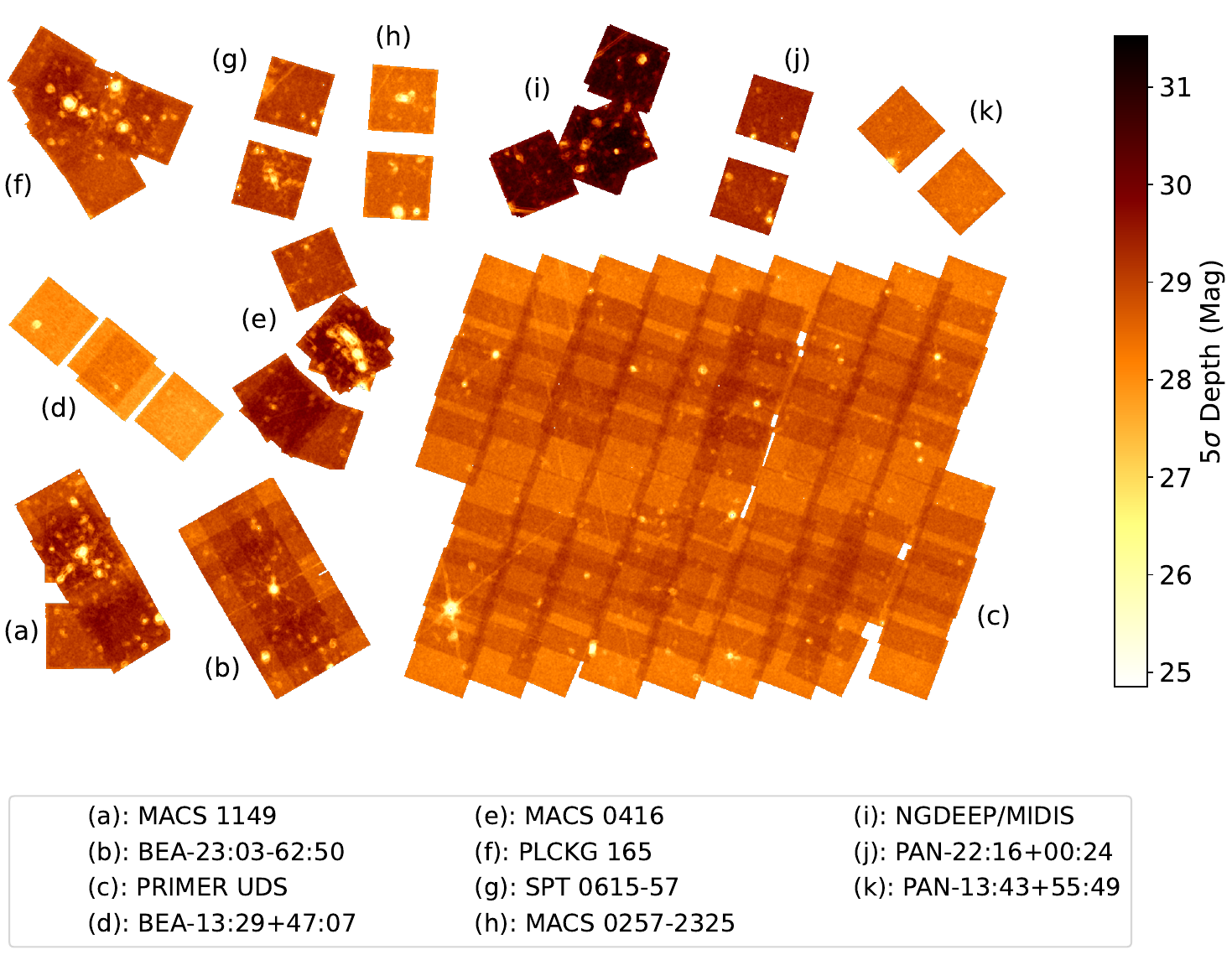}
    \caption{Depth maps corresponding to 5$\sigma$ limiting magnitudes ($0.20^{\prime\prime}-$diameter apertures and point-source corrected to total) for a small subset of the fields studied. This subset spans $\simeq10\%$ of the area and $\simeq7\%$ of the independent sightlines, illustrating the dynamic range in sensitivity and area that this study has probed. The fields are scaled to reflect their relative sizes.}
    \label{fig:Depth_Maps}
\end{figure*}

The global 5$\sigma$ limiting magnitudes for each NIRCam dataset are provided in Appendix A, Tables \ref{table: depths}-\ref{table: depths3}. These were calculated by measuring the absolute median deviation (MAD) of blank-sky $0.20^{\prime\prime}-$diameter apertures spanning the field-of-view, scaling to $\sigma=1.4826\times$MAD and then correcting to total assuming a point-source correction. In Fig. \ref{fig:Areas_Depths}, we show the cumulative distribution of 8$\sigma$ global depth in F277W (or F250M where F277W is unavailable) plotted against survey area. Notably, $>70\%$ of our area reaches a $8\sigma$ limiting AB magnitude $> 28.0$ (0.20$^{\prime\prime}$--diameter aperture, corrected to point-source total), providing a powerful overall combination of depth and area. In Fig. \ref{fig:Depth_Maps} we show $5\sigma$ depth maps for a small subset of our data set to further illustrate the dynamic range in data probed. We also note that within the overall raw survey area probed, there is an approximately $\lesssim2\%$ area overlap, split between fifteen different pointings, typically where some modest fraction of a pure parallel pointing overlaps with a legacy field e.g. PRIMER, CEERS, or JADES.

\section{SAMPLE SELECTION}
\label{section3}
\subsection{Catalogue construction}
For each NIRCam field, we constructed multi-wavelength catalogues by running \textsc{Source-Extractor} \citep{Bertin1996} in dual-image mode. For the vast majority of fields, this was done with F200W, F277W and F356W detection images. However, for a very small subset of fields, we used F210M and/or F250M detection images in the absence of F200W and/or F277W. In order to maximise the sensitivity to ultra high-redshift galaxies, we adopted 0.20$^{\prime\prime}-$diameter apertures. Aperture photometry was measured on imaging that had been PSF-homogenized to match the resolution of the F444W imaging. Photometric uncertainties were measured by taking the aperture-to-aperture rms of the nearest 200 blank-sky apertures, after removing sources using a segmentation map which was dilated by an additional two pixels. In order to account for any remaining differences in the PSF curve of growth between bands, we applied aperture corrections (typically at the $\simeq1-2\%$ level) to the fluxes and uncertainties.

We imposed signal-to-noise cuts on the initial catalogues prior to spectral energy distribution (SED) fitting. Firstly, as we demand no significant detection of flux shortward of $\lambda\simeq1.64\,\mu m$ (i.e. the Lyman break at $z=12.5$), we required $\leq2\sigma$ non-detections in NIRCam F090W and F115W, as well as in F140M where available. Although the main objective of this study is to measure the galaxy UV LF over the redshift range $12.5<z<18.5$ with a sample of F150W-dropouts, it is actually not until $z\geq12.7$ that the Lyman break completely leaves the F150W filter coverage; at $z=12.5$ there is still $\lesssim10\%$ transmission in F150W, which can allow for especially bright candidates to possibly yield a F150W detection. We accounted for this by imposing a $\leq2\sigma$ non-detection criteria for F150W over the entire redshift window of interest except for the small redshift window $12.4\leq z\leq 12.7$, where we required this F150W $\leq2\sigma$ non-detection criteria \textit{or} a $F150W-F200W\geq1.5$ colour. This gives similar red break colours to a $2\sigma$ F150W and $8\sigma$ F200W high-redshift candidate in the case of F150W and F200W being of the same depth, but still allows for some marginal detection in F150W for bright sources at the very edge of the F150W filter transmission curve. A notable example of this is the spectroscopically confirmed GHZ2 at $z_{\rm spec}=12.34$ \citep{castellano2024}, which has a $z_{\rm phot}=12.5\pm0.1$, is  detected at SNR$\simeq5$ in F150W, but has a $>2.5$ mag $F150W-F200W$ Lyman-break colour.

Where we include {\it HST} ACS imaging, we also require a $\leq3\sigma$ non-detection, imposing this slightly more permissive non-detection criterion to allow for the possibility of non-Gaussian noise in the noisier \textit{HST} imaging.

We then required either a F200W or F277W detection at $\geq8\sigma$ significance, in order to ensure that we are dealing with robustly-detected sources. The primary motivation for a more conservative detection threshold versus, e.g., 5$\sigma$, was to ensure that we are able to robustly define the Lyman-break spectral discontinuity between F150W and F200W, even in cases where F150W is $\simeq 0.5$\,mag. shallower than the longer-wavelength imaging in a given data set. In our previous work, \citet{donnan2024}, we demonstrated that when pushing to the $5\sigma$ limit, one requires the use of an LF prior to rule out potential contamination by low-redshift interlopers. However, with the conservative $8\sigma$ threshold adopted in this study, and in the \citep{mcleod2024} study at $z=11$, the resulting strength of the Lyman break removes the need for the prior. We note that, at $z=11$, there is excellent agreement between the results of \citet{mcleod2024} and \citet{donnan2024}, despite the different approaches to sample selection.

\subsection{SED fitting}
Next, we ran the SED fitting code \textsc{LePhare} \citep{Arnouts1999} in order to measure photometric redshifts, utilising the \citet{BC03} models and a \citet{Chabrier2003} initial mass function (IMF), accounting for the IGM absorption as in \citet{Madau1995}. Included are declining star-formation histories with $\tau$ ranging from 0.1 to 15 Gyr, a \citet{Calzetti2000} dust attenuation law with allowed $A_{V}$ values [0, 6.0, 0.2], and two metallicities: 0.2$Z_{\odot}$ and $Z_{\odot}$. We also included emission lines in our fits.

We measured the rest-frame UV magnitude ($M_{1500}$) for all objects using a 100\,\AA\,top-hat filter centered on 1500\,\AA\,on the best-fitting SED. To correct this to total, we followed our previous work in \citet{mcleod2024} and \citet{cullen2024}, and scaled our measurements to the corresponding \textsc{Source Extractor} \textsc{FLUX\_AUTO} measurement, with the addition of a further $\simeq10\%$ to account for light outside the Kron aperture \citep{Kron1980}.

At this stage, we retained all objects with best-fitting solutions $z_{\rm phot}\geq11.5$ and $\Delta \chi^{2}\geq 4$, which we then subjected to rigorous visual inspection in order to clean the sample for artefacts as well as verifying the non-detections at $\lambda\leq1.5\,\mu m$ on the imaging. We also confirmed the non-detections within stacks of the available short-wavelength NIRCam imaging shortward of, and including, the F150W filter.

\section{High-redshift candidates}
\label{section4}
\begin{figure*}
    \centering
    \begin{tabular}{cc}
    \includegraphics[width=0.45\textwidth]{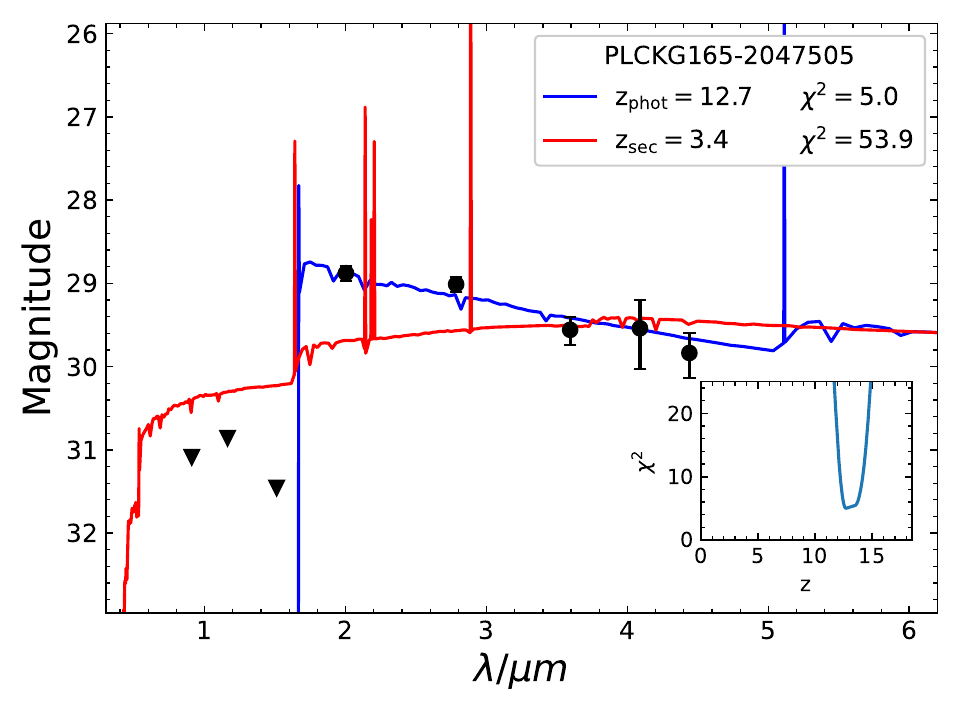} &
    \includegraphics[width=0.45\textwidth]{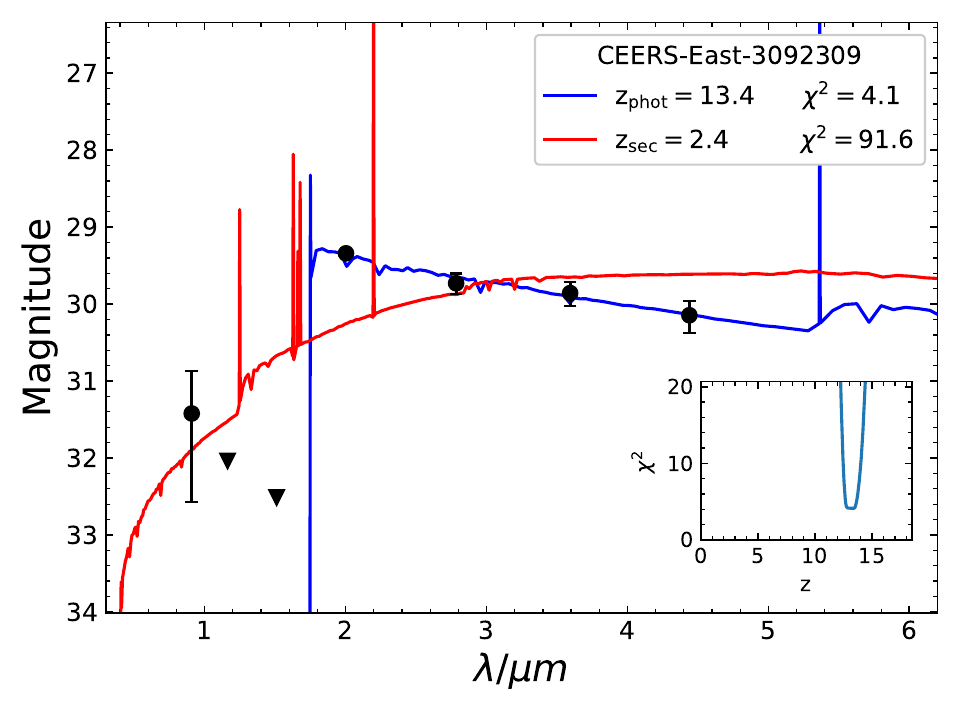}  \\[-3ex]
    \includegraphics[width=0.45\textwidth]{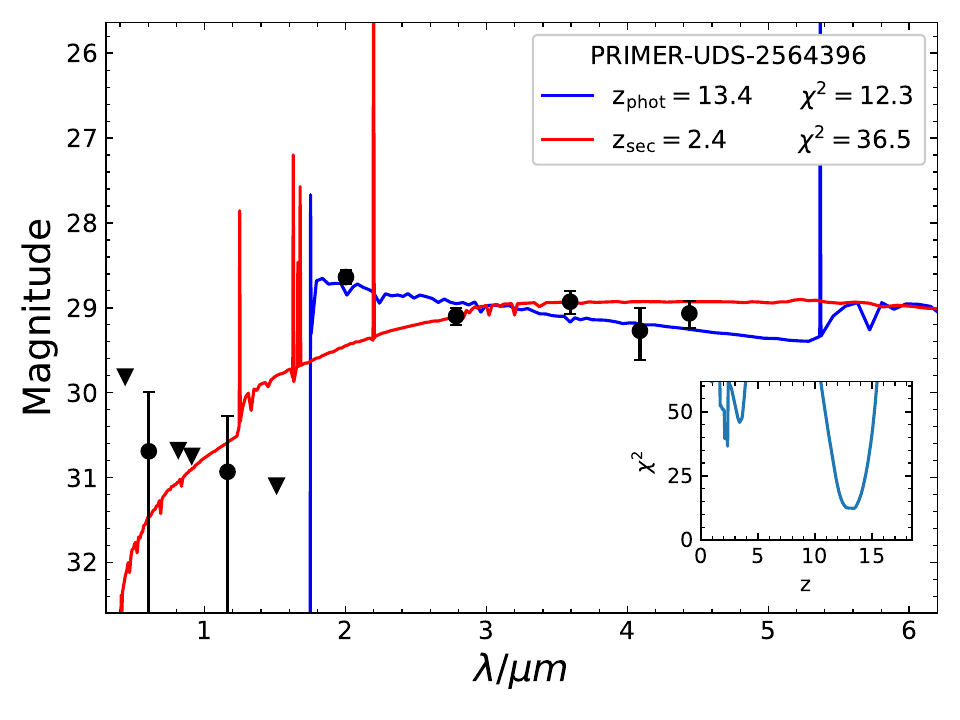}  &
    \includegraphics[width=0.45\textwidth]{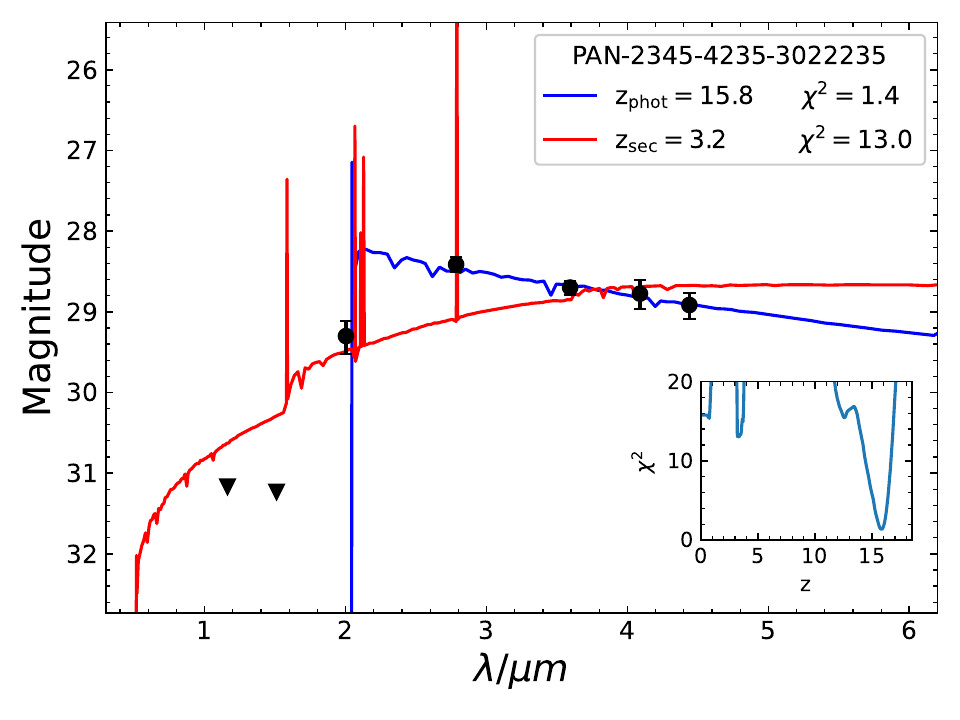} \\[-3ex]
    \end{tabular}
    \caption{Example SED plots, including an inset $\chi^{2}$ distribution, for a subset of our high-redshift candidates. The photometry presented is in 0.20$^{\prime\prime}$--diameter apertures. For each source we show both the primary photometric redshift solution in blue and the (disfavoured) secondary low-redshift solution in red. For non-detections we indicate 1$\sigma$ limiting magnitudes with downward arrow symbols.}
    \label{fig: SEDs}
\end{figure*}

\begin{figure*}
    \centering
    \includegraphics[width=0.9\textwidth]{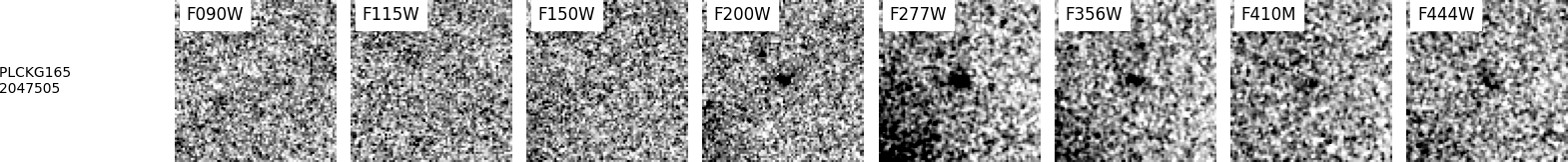} \\
    \includegraphics[width=0.9\textwidth]{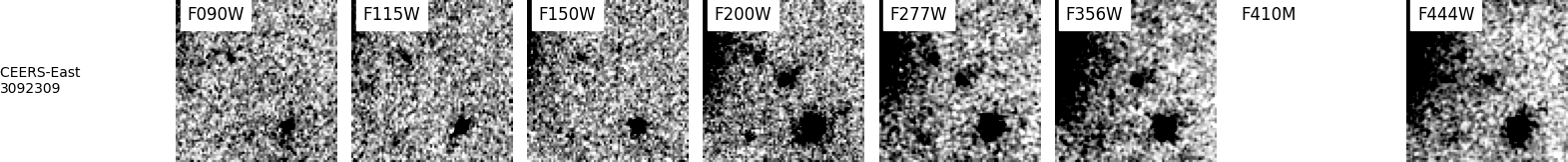}  \\
    \includegraphics[width=0.9\textwidth]{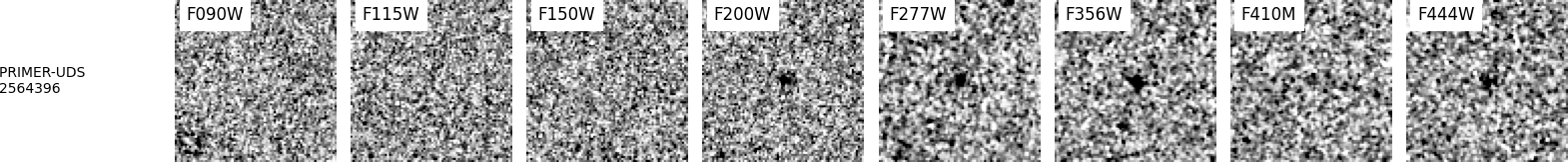}  \\
    \includegraphics[width=0.9\textwidth]{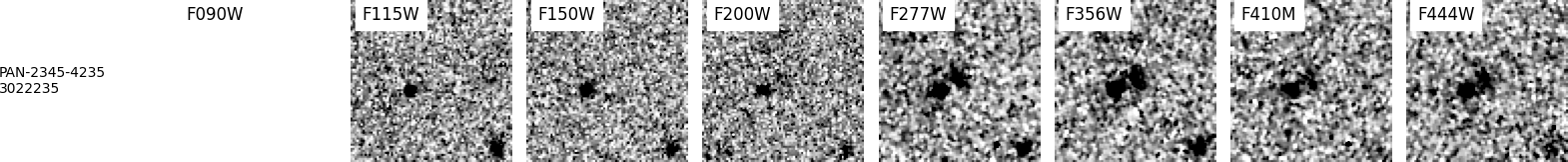} \\
    \caption{Example NIRCam postage stamps (from left to right: F090W, F115W, F150W, F200W, F277W, F356W, F410M, F444W where available) for the same subset of our high-redshift candidates as in Fig \ref{fig: SEDs} (North=up, East=left). The scaling is the median $\pm\,2\sigma$ after masking objects.}
    \label{fig: Stamps}
\end{figure*}

The final $z\geq11.5$ galaxy candidate list is given in Tables\,\ref{candidate_list1}-\ref{candidate_list2}. We include SED fits and postage stamps for an example subset of our selected objects in Figs\,\ref{fig: SEDs}-\ref{fig: Stamps}. Although, for completeness, we report any $11.5<z<12.5$ candidates uncovered in this search, our insistence on F150W non-detection (or strong red $\mathrm{F150W-F200W}\geq1.5$ colour) precludes any robust statistical study of the $11.5<z<12.5$ galaxy population here.

A number of recent studies have reported evidence for significant populations of galaxies at $z>12$ \citep{mcleod2024, finkelstein2024, conselice2024}. While it is generally the case that the early \textit{JWST} studies were in rather good agreement on the overall shape and evolution of the UV LF between $z \simeq 8$ and $z\simeq 11$, there were considerable discrepancies in the individual candidates reported \citep{bouwens2023}. Some of this was undoubtedly due to differences in early NIRCam reduction techniques (e.g., flux calibrations, pipeline version, or CRDS context). Other differences can be quite reasonably attributed to different selection methods/criteria (e.g., colour-selection versus full SED fitting, or different adopted limiting magnitude and non-detection thresholds). The discrepancies in some of the early high-redshift candidate lists helped motivate our decision to adopt a relatively conservative SNR$\geq8\sigma$ cut in the candidate selection process utilized here, as previously also applied by \citet{mcleod2024}. As is to be expected, comparison between the $z\simeq10-11$ candidates identified by \citet{mcleod2024} and those identified in previous studies, such as \citet{Adams2024}, \citet{finkelstein2024} and \citet{castellano2024}, confirms that our insistence on $\geq8\sigma$ selection produces smaller, but more robust and repeatable candidate lists.

\subsection{$\mathbf{z\geq15}$ candidates uncovered in this study}

From our wide-area search of the PANORAMIC data, we identified three credible $z\geq15$ candidates. One apparently robust $z\simeq15$ candidate was PAN-z14-1, at RA 22:17:00.11, Dec +00:22:44.69. This ultra high-redshift candidate was found in one of the PANORAMIC pure-parallel pointings in the region near SSA22, which is a well-studied $z\simeq3$ protocluster (e.g., \citealt{topping2018}). It was the subject of a successful proposal for follow-up with \textit{JWST} NIRSpec single-slit observations in Cycle 4 (PID 6954; PIs Donnan, McLeod). The object was subsequently also found by \citet{Weibel2025}.

While the NIRSpec spectroscopy of PAN-z14-1 confirmed its high-redshift nature, its true redshift was found to be $z_{\rm spec}=13.5$ \citep{donnan2026}, somewhat lower than inferred from the photometry (in part due to the impact of the Lyman-damping wing on the inferred position of the Lyman break). Although ultimately not a redshift record breaker, this galaxy is one of the the brightest spectroscopically confirmed sources at $z>12$ ($M_{\rm UV}=-20.6$) and is thus located at the very bright end of the $z=13.5$ UV LF.

The two other apparently robust $z > 15$ candidates that we uncovered within the PANORAMIC survey have yet to be scrutinized with NIRSpec. At $z_{\rm phot}=15.8\pm0.3$, PAN-2345-4234-3022235 (PAN-z16-1) is the highest redshift galaxy candidate we uncovered. It is also remarkably bright, $M_{1500}=-20.9$ and very blue $\beta\simeq-3.0$ (Bondestam et al. 2026, in prep). The other candidate, PAN-0332-2753-2030538, with $z_{\rm phot}=15.7^{+0.5}_{-0.4}$, is in the GOODS-South region, and was previously reported by \citet{hainline2026} at a similarly high redshift of $z_{\rm phot}\simeq16$ (see Section  \ref{section: comparison_candidates}).

\subsection{PLCKG165-2047505}
\label{section:plckg165}
When combining several epochs of imaging covering the PLCKG165 cluster \citep{frye2024}, we selected a notable robust $z\simeq13$ source, PLCKG165-2047505, at 11:27:13.98, +42:28:56.16. With a magnification of $\mu=3.2$, derived using the \citet{Kamieneski2024} lensing model, this is the most strongly magnified source in our sample. This galaxy candidate will be followed up with NIRSpec as part of the \textit{JWST} Cycle 5 PID 11171 (PIs McLeod, Garuda).

\begin{table*}
\begin{tabular}{lccccccl}
\hline
ID & RA & Dec & $\rm z_{\rm phot}$ & $M_{1500}$ & $\rm \mu$ & z$_{\rm spec}$ & Reference\\
\hline
2236+0032-2027652 & 22:36:44.57 & $+$00:32:18.57 & $11.6^{+0.7}_{-1.0}$ & $-20.0$ & - & - & - \\[1ex]
PAN-0134-1531-2023208 & 01:34:43.79 & $-$15:31:27.39 & $11.8^{+0.6}_{-0.7}$ & $-20.5$ & - & - & - \\[1ex]
PAN-1707+5852-2015557 & 17:07:15.91 & $+$58:51:44.10 & $11.9^{+0.5}_{-0.5}$ & $-20.4$ & - & - & - \\[1ex]
PAN-0104-5508-2026296 & 01:04:17.50 & $-$55:07:35.55 & $12.0^{+0.4}_{-0.5}$ & $-20.7$ & - & - & - \\[1ex]
PAN-1007+2109-2040861 & 10:07:44.02 & $+$21:11:00.98 & $12.0^{+0.4}_{-0.3}$ & $-20.6$ & - & - & - \\[1ex]
UNCOVER-2079090 & 00:14:29.18 & $-$30:27:22.84 & $12.1^{+0.4}_{-0.4}$ & $-19.0$ & 1.2 & - & - \\[1ex]
PAN-1343+5549-2000350 & 13:43:39.91 & $+$55:46:45.76 & $12.1^{+0.5}_{-0.4}$ & $-21.2$ & - & - & - \\[1ex]
PAN-1707+5852-2010068 & 17:07:17.36 & $+$58:51:11.06 & $12.2^{+0.5}_{-0.6}$ & $-21.4$ & - & - & - \\[1ex]
PRIMER-COS-2274136 & 14:19:12.35 & $+$52:51:41.81 & $12.2^{+0.5}_{-0.4}$ & $-18.8$ & - & - & - \\[1ex]
J1235-2051316 & 12:35:53.39 & $+$04:55:53.04 & $12.2^{+1.5}_{-0.5}$ & $-18.7$ & - & - & - \\[1ex]
CEERS-East-1088018 & 14:20:12.28 & $+$52:48:32.66 & $12.3^{+0.5}_{-0.4}$ & $-18.3$ & - & - & - \\[1ex]
CEERS-2264171 & 14:19:27.31 & $+$52:51:29.20 & $12.3^{+0.4}_{-0.4}$ & $-19.3$ & - & - & - \\[1ex]
JADES-GS-2104976 & 03:32:25.12 & $-$27:51:42.74 & $12.3^{+0.3}_{-0.3}$ & $-17.9$ & - & - & - \\[1ex]
NEP-TDF-2034945 & 17:22:57.48 & $+$65:46:29.09 & $12.3^{+0.4}_{-0.3}$ & $-19.5$ & - & - & - \\[1ex]
JADES-GS-1061386 & 03:32:20.33 & $-$27:52:00.02 & $12.3^{+0.4}_{-0.5}$ & $-18.0$ & - & - & - \\[1ex]
JADES-GN-3145713 & 12:37:30.39 & $+$62:15:15.68 & $12.4^{+0.3}_{-0.4}$ & $-19.8$ & - & - & - \\[1ex]
JADES-GS-2199196 & 03:32:39.92 & $-$27:49:17.62 & $12.4^{+0.1}_{-0.2}$ & $-19.0$ & - & 12.63 & \citet{Curtis-Lake2023} \\[1ex]
JADES-GS-2037358 & 03:32:06.88 & $-$27:53:34.85 & $12.4^{+0.3}_{-0.3}$ & $-18.5$ & - & - & - \\[1ex]
PAN-1237+6210-1016125 & 12:37:46.16 & $+$62:11:49.72 & $12.4^{+2.6}_{-0.5}$ & $-19.4$ & - & - & - \\[1ex]
CEERS-East-1160411 & 14:20:03.15 & $+$52:52:30.49 & $12.4^{+0.4}_{-0.4}$ & $-19.1$ & - & - & - \\[1ex]
BEA-2303-6250-2039802 & 23:03:34.42 & $-$62:50:11.26 & $12.4^{+1.3}_{-0.3}$ & $-21.0$ & - & - & - \\[1ex]
CEERS-East-3024423 & 14:19:49.00 & $+$52:44:06.57 & $12.4^{+0.2}_{-0.3}$ & $-19.4$ & - & - & - \\[1ex]
BEA-0217-0503-2052739 & 02:17:13.11 & $-$05:01:50.54 & $12.5^{+1.4}_{-0.4}$ & $-19.0$ & - & - & - \\[1ex]
GLASS-2039052 & 00:13:59.75 & $-$30:19:29.14 & $12.5^{+0.1}_{-0.1}$ & $-21.0$ & 1.2 & 12.34 & \citet{castellano2024} \\[1ex]
PAN-1000+0207-2032751 & 10:00:24.17 & $+$02:08:31.01 & $12.5^{+1.7}_{-0.4}$ & $-20.1$ & - & - & - \\[1ex]
BEA-2303-6250-1006779 & 23:03:57.79 & $-$62:52:17.48 & $12.6^{+1.3}_{-0.5}$ & $-19.1$ & - & - & - \\[1ex]
GLASS-2014204 & 00:14:03.26 & $-$30:21:24.40 & $12.6^{+1.6}_{-0.2}$ & $-19.1$ & 1.7 & - & - \\[1ex]
PAN-0144+1714-2042105 & 01:44:22.44 & $+$17:15:31.36 & $12.6^{+1.0}_{-0.3}$ & $-19.1$ & - & - & - \\[1ex]
JADES-GS-1074873 & 03:32:05.31 & $-$27:51:26.05 & $12.6^{+1.7}_{-0.4}$ & $-19.2$ & - & - & - \\[1ex]
BEA-2303-6250-2027277 & 23:03:47.47 & $-$62:51:07.34 & $12.6^{+1.2}_{-0.3}$ & $-19.9$ & - & - & - \\[1ex]
SPT0615-1020236 & 06:15:41.93 & $-$57:43:59.79 & $12.7^{+1.1}_{-0.5}$ & $-19.6$ & 1.2 & - & - \\[1ex]
MACS0416-1003914 & 04:16:16.35 & $-$24:07:41.72 & $12.7^{+0.9}_{-0.4}$ & $-19.2$ & - & - & - \\[1ex]
NGDEEPMIDIS-2043718 & 03:32:58.43 & $-$27:51:02.90 & $12.7^{+0.7}_{-0.4}$ & $-17.6$ & - & - & - \\[1ex]
PLCKG165-2047505 & 11:27:13.98 & $+$42:28:56.16 & $12.7^{+1.0}_{-0.2}$ & $-18.6$ & 3.2 & - & - \\[1ex]
BEA-0446-2636-2034805 & 04:46:42.76 & $-$26:34:42.75 & $12.7^{+1.5}_{-0.4}$ & $-19.8$ & - & - & - \\[1ex]
NGDEEPMIDIS-2051940 & 03:33:02.18 & $-$27:50:20.12 & $12.8^{+0.8}_{-0.3}$ & $-17.3$ & - & - & - \\[1ex]
PRIMER-UDS-2138854 & 02:17:01.80 & $-$05:17:18.67 & $12.9^{+0.8}_{-0.3}$ & $-19.6$ & - & - & - \\[1ex]
BEA-2058-4247-2067175 & 20:58:33.74 & $-$42:44:08.61 & $13.0^{+0.9}_{-0.4}$ & $-21.0$ & - & - & - \\[1ex]
JADES-GS-2042986 & 03:32:15.54 & $-$27:53:24.86 & $13.0^{+0.9}_{-0.5}$ & $-18.4$ & - & 13.01 & \citet{witstok2025} \\[1ex]
NGDEEPMIDIS-2051280 & 03:33:00.74 & $-$27:50:21.41 & $13.2^{+0.4}_{-0.7}$ & $-18.0$ & - & - & - \\[1ex]
NGDEEPMIDIS-2041809 & 03:32:56.97 & $-$27:51:10.65 & $13.2^{+0.3}_{-0.5}$ & $-17.5$ & - & - & - \\[1ex]
CEERS-East-3092309 & 14:20:06.05 & $+$52:48:15.18 & $13.4^{+0.3}_{-0.7}$ & $-19.2$ & - & - & - \\[1ex]
PRIMER-UDS-2564396 & 02:17:10.08 & $-$05:10:26.89 & $13.4^{+0.2}_{-0.8}$ & $-20.1$ & - & - & - \\[1ex]
UNCOVER-2192918 & 00:14:12.63 & $-$30:21:00.76 & $13.4^{+0.7}_{-1.0}$ & $-18.0$ & 1.9 & - & - \\[1ex]
\hline
\end{tabular}
\caption{Basic information for the final sample of $11.5\leq z<13.5$ galaxies across all of the fields studied. The absolute UV magnitude $M_{1500}$ is corrected to total and de-lensed by the magnification $\mu$ where applicable. Note that a $\mu=1.2$ floor is applied when a candidate is in a lensed field, but lies outside of the lensing map coverage. For any object which has a spectroscopic redshift reported in a previous study, we provide a relevant reference in the final column.}
\label{candidate_list1}
\end{table*}

\begin{table*}
\begin{tabular}{lccccccl}
\hline
ID & RA & Dec & $\rm z_{\rm phot}$ & $M_{1500}$ & $\rm \mu$ & z$_{\rm spec}$ & Reference\\
\hline
NGDEEPMIDIS-2048620 & 03:32:58.20 & $-$27:50:38.72 & $13.5^{+0.2}_{-1.1}$ & $-17.7$ & - & - & - \\[1ex]
MACS0416-1003912 & 04:16:16.33 & $-$24:07:41.72 & $13.5^{+0.2}_{-0.3}$ & $-18.9$ & - & - & - \\[1ex]
JADES-GS-2084600 & 03:32:11.05 & $-$27:52:10.66 & $13.6^{+0.6}_{-1.4}$ & $-18.9$ & - & - & - \\[1ex]
JADES-GN-2285047 & 12:36:40.16 & $+$62:18:36.94 & $13.9^{+0.5}_{-0.5}$ & $-19.5$ & - & - & - \\[1ex]
J1235-2062314 & 12:35:55.12 & $+$04:56:51.50 & $14.1^{+0.4}_{-1.8}$ & $-19.5$ & - & - & - \\[1ex]
CEERS-2507584 & 14:20:17.83 & $+$52:57:04.12 & $14.2^{+0.4}_{-1.7}$ & $-19.4$ & - & - & - \\[1ex]
JADES-GS-2051088 & 03:32:17.83 & $-$27:53:09.30 & $14.4^{+0.7}_{-2.1}$ & $-19.1$ & - & 13.90 & \citet{carniani2024} \\[1ex]
PAN-2216+0024-3003658 & 22:17:00.12 & $+$00:22:44.69 & $15.0^{+0.2}_{-0.3}$ & $-20.6$ & - & 13.53 & \citet{donnan2026} \\[1ex]
PRIMER-COS-2332181 & 10:00:22.40 & $+$02:16:23.35 & $15.0^{+0.4}_{-0.3}$ & $-20.0$ & - & 14.44 & \citet{naidu2025} \\[1ex]
PAN-0332-2754-2030538 & 03:32:41.21 & $-$27:53:45.00 & $15.7^{+0.5}_{-0.4}$ & $-19.1$ & - & - & - \\[1ex]
PAN-2345-4235-3022235 & 23:45:15.45 & $-$42:34:57.07 & $15.8^{+0.3}_{-0.3}$ & $-20.9$ & - & - & - \\[1ex]
\hline
\end{tabular}
\caption{Basic information for the final sample of $z\geq13.5$ galaxies across all of the fields studied. The absolute UV magnitude $M_{1500}$ is corrected to total and de-lensed by the magnification $\mu$ where applicable. For any object which has a spectroscopic redshift reported in a previous study, we provide a relevant reference in the final column.}
\label{candidate_list2}
\end{table*}

\section{The galaxy UV LUMINOSITY FUNCTION}
\label{section5}
\subsection{Effective volumes and completeness}
To derive the evolving galaxy luminosity function, we employed the $1/V_{\rm max}$ method \citep{Schmidt1968}. As detailed above, a key feature of this study is the use of a large number of non-contiguous lines-of-sight. Although this mitigates the effects of cosmic variance, it inevitably required us to deal carefully with a heterogeneous range of depths across the many different survey fields.

We defined the overall {\it effective volumes} object-by-object, and field-by-field. To do this, we first constructed F277W depth maps for each of our fields, taking each pixel and measuring the aperture-to-aperture rms at that location from the nearest 200 blank-sky apertures (in an identical fashion to our photometric uncertainty measurements). We illustrate a selection of the depth maps, designed to demonstrate the range of depths in our overall search, in Fig. \ref{fig:Depth_Maps}. As described above, our final galaxy sample comprises those objects for which $\mathrm{F277W\geq8\sigma}$ in the field where they were found. For each galaxy, we therefore calculated an effective search area by summing up the contribution of all pixels, across all fields, that would have been sufficiently deep to detect it at 8$\sigma$ significance in F277W. For this area we then calculated the maximum redshift ($z_{\rm max}$), and hence the associated maximum volume ($V_{\rm max}$).

For any lensed candidates, we de-lensed their observed flux density, considering only their intrinsic flux density when assigning their effective volume in the overall survey. For the lensed fields, we therefore de-lensed both the survey depths and effective areas to account for the ability of these fields to reveal intrinsically fainter objects, but in a smaller cosmological volume. For areas of cluster fields not covered by the magnification map (i.e., the parallel NIRCam module), we assigned a magnification floor, $\mu$, which was typically $\mu=1.2$. For a small number of fields where the $\mu$ map was already $\mu<1.2$ at the edge, this floor was adjusted down to $\mu=1.1$ or $\mu=1.0$ as appropriate. In practice, our results are insensitive to the choice of this $\mu$ floor, as such regions constitute a very modest fraction ($\simeq5-10\%$) of the total survey area probed.

For each of our fields, we estimated the completeness as a function of the F277W magnitude via a series of source injection and recovery simulations. We ran these simulations over the magnitude range $m_{F277W}=25-31$ in steps of 0.05\,mag. At each simulation step, we injected 1000 simulated high-redshift galaxies, and ran each full simulation 30 times. For all synthetic galaxies, we assumed point-source profiles. We ensured that the synthetic galaxies also had the expected non-detections at SNR$<$2 in the {\it JWST} NIRCam short-wavelength filters and SNR$<3$ in the {\it HST} ACS filters. Based on this analysis, for each of our high-redshift candidates, we corrected the effective volume contribution field-by-field based on the completeness measured for their specific F277W magnitude.

To calculate the uncertainties for our UV LF number densities, we employed a bootstrap analysis, computing $\sigma_{\rm boot}$ by sampling 50,000 realisations of the UV LF. We then combined these in quadrature with the estimated cosmic variance uncertainties, $\sigma_{\rm CV}$. The cosmic variance uncertainty is estimated in the following way. We first consider for each object how many fields contribute to the $V_{\rm max}$ measurement, and thus average this across all of the objects in a magnitude bin to get N. We then employ the \citet{Trenti2008} cosmic variance calculator to estimate the fractional cosmic variance uncertainty for a single NIRCam pointing at $z=13.5$, the centre of the LF bin, conservatively assuming that every sight-line is a single NIRCam pointing. We then scale down this value by $\sqrt N$ to determine the fractional cosmic variance uncertainty for the bin, which we convert to $\sigma_{\rm CV}$. In practice, the cosmic variance uncertainty contribution is rendered negligible across all bins ($\simeq5\%$ level) thanks to the number of sightlines probed. We present the number densities for each of our UV LF bins in Table \ref{tab:number_densities}.

Although the gravitationally-lensed regions represent a modest fraction of our overall survey area, it is instructive to test how sensitive our UV LF results are to the inclusion of highly lensed regions. We therefore perform two tests whereby we restrict our search area to only regions of modest magnification, first $\mu<5$, and then $\mu<3$. Masking the $\mu>5$ regions results in a near-negligible change in our overall LF, with $\simeq0-2\%$ higher number densities at bright magnitudes, and $\simeq3-6\%$ higher at the faint end. For the more extreme case of restricting to only the modest magnification regions, $\mu<3$, we still find a minimal increase on the bright-end of the LF ($M_{1500}<M_{1500}^{\star}$) at the $\simeq1-2\%$ level, and a $\simeq3-9\%$ change at fainter magnitudes. We can hence be confident that our results are insensitive to the impact of gravitational lensing.

\begin{table}
    \centering
    \begin{tabular}{ccc}
    \hline
       z & $M_{1500}$ & $\phi$ / $10^{-6}$ Mpc$^{-3}$ mag$^{-1}$\\
        \hline
     \multirow{7.5}{*}{$12.5<z<14.5$}  & $-$21.700\phantom{*} & \phantom{*}$<$\,0.3 \\[1ex]
       & $-$20.825\phantom{*} & \phantom{*}0.8 $^{+0.4}_{-0.4}$ \\[1ex]
       & $-$20.075\phantom{*} & \phantom{*}2.8 $^{+1.4}_{-1.2}$ \\[1ex]
       & $-$19.450\phantom{*} & 14.8 $^{+8.6}_{-7.1}$ \\[1ex]
        & $-$18.950\phantom{*} & 37.6 $^{+18.6}_{-15.8}$ \\[1ex]
        & $-$18.325\phantom{*} & 92.6 $^{+59.1}_{-46.4}$ \\[1ex]
    \hline
    \multirow{5}{*}{$14.5<z<18.5$} &   $-$21.000\phantom{*} & \phantom{*}0.19 $^{+0.19}_{-0.19}$ \\[1ex]
    & $-$20.000\phantom{*} & $<$\,0.23 \\[1ex]
    & $-$19.000\phantom{*} & \phantom{*}0.84 $^{+0.84}_{-0.84}$ \\[1ex]
    & $-$18.000\phantom{*} & $<$\,12 \\[1ex]
    \hline
    \end{tabular}
    \caption{The number densities measured for both the $z=13.5$ LF (top) and the $z=15.8$ LF (bottom).}
    \label{tab:number_densities}
\end{table}

\subsection{The UV LF at $\mathbf{z\simeq13.5}$}
\begin{figure*}
    \begin{tabular}{cc}
    \centering
    \includegraphics[width=0.5\textwidth]{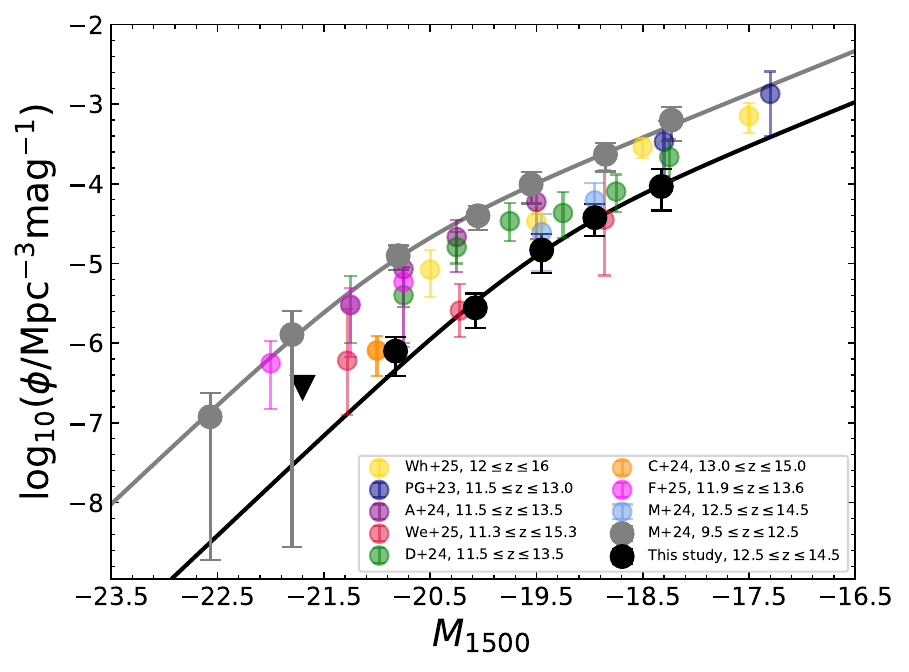} &
    \includegraphics[width=0.5\textwidth]{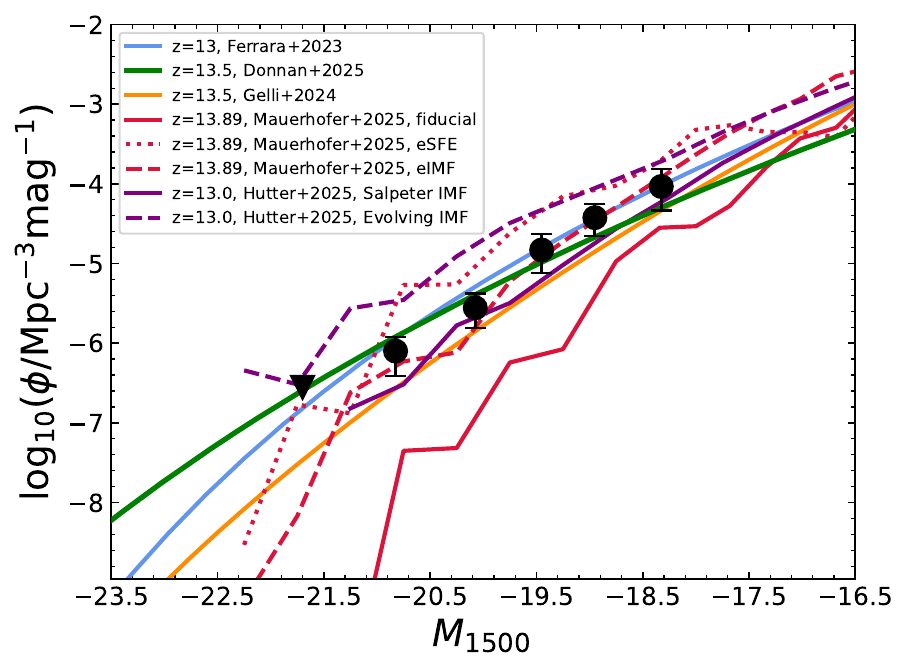}
    \end{tabular}
    \caption{Our new determination of the galaxy UV luminosity function at $12.5<z<14.5$. In the left-hand panel, we include, for comparison, the  $z\simeq11$ UV LF determination from \citet{mcleod2024}, as well as a number of other observational studies at similar redshifts. In the right-hand panel, we compare our results with the predictions of several recent theoretical models of galaxy evolution. The full references corresponding to the literature results indicated in the left-hand panel legend are: Wh+25=\citet{whitler2025}; PG+23=\citet{Perez-Gonzalez2023}; A+24=\citet{Adams2024}; We+25=\citet{Weibel2025}; D+24=\citet{donnan2024}; C+24=\citet{casey2024}; F+25=\citet{Franco2025} M+24=\citet{mcleod2024}.}
    \label{fig: UV_LF_z13p5}
\end{figure*}
In Fig. \ref{fig: UV_LF_z13p5} we present our determination of the evolving UV LF at $12.5\leq z \leq14.5$, and compare this with the $z\simeq11$ UV LF from \citet{mcleod2024} as well as a compendium of results from the recent literature at similar redshifts. We are able to constrain our UV LF number densities over a broad dynamic range in $M_{1500}$, spanning $\simeq3$ mag. At fainter magnitudes, there is agreement within $1\sigma$ for the number density bins overlapping with our previous, tentative results at $12.5\leq z\leq14.5$. However, at the bright end, i.e. brighter than the characteristic magnitude ($M^{\star}_{1500}=-20.87$) as found for the double power-law fit at $z\simeq11$ from \citet{mcleod2024}, we find a relative lack of galaxies over the higher-redshift range, with only one galaxy at $M_{1500}\simeq-21.0$. We hence find evidence of a strong decline in the number densities of galaxies between our previous $9.5\leq z \leq 12.5$ UV LF \citep{mcleod2024} and the present study, with the evolution being more pronounced at brighter luminosities.

We fit a double power-law function to our data, fixing both the bright-end and faint-end slopes to our previous \citet{mcleod2024} measurements at $z=11$. This fiducial fit is shown in Fig \ref{fig: UV_LF_z13p5}, and the parameters are provided in Table \ref{tab:params}. Alternatively, adopting a double power-law fit and assuming a pure $M^{\star}$ evolution between our $z=11$ and $z=13.5$ measurements, we find that $M^{\star}$ dims by $\simeq1.3$ mag. If instead we assume a pure $\phi^{\star}$ evolution in the double power-law fit, $\phi^{\star}$ drops by a factor of $\simeq12$ with respect to our $z\simeq11$ measurement. We note that the pure luminosity evolution is a much better match to the data versus the pure $\phi^{\star}$ evolution, with the latter fit barely able to match within $\simeq1\sigma$ of each of the datapoints. The number densities at $M_{\rm UV}\leq-20$ also suggest a clear departure from the gradual evolution of the bright-end of the LF seen over the redshift range $9<z<12$ (as now well established by a series of earlier {\it JWST} studies: e.g., \citealt{Bowler2020,Donnan2023a,mcleod2024,donnan2024}).

\subsection{A consistent picture of the UV LF at extreme redshifts?}
Our $z \simeq 13.5$ UV LF is in good agreement with some other published studies at comparable redshifts. In particular, although they probe a broader redshift range $11.3<z<15.3$, our number density measurements are very similar to those of \citet{Weibel2025} over the common range in $M_{1500}$. This is perhaps not completely surprising, given that the \citet{Weibel2025} study also probes a wide area, the studies share a significant overlap, and they also employ conservative $8\sigma$ detection thresholds. The \citet{Weibel2025} work also benefits from reduced field-to-field variance from the large number of non-contiguous pointings.

We are also in agreement with the $13<z<15$ measurement in a single-bin at $M_{\rm UV}=-21$ from COSMOS-Web \citep{casey2024}. At first glance, this is encouraging, given that they also probe a fairly large area of $\simeq0.28$\,deg$^2$ but mostly covering a different region of the sky (the exceptions being the PRIMER COSMOS footprint, and the overlapping regions between COSMOS-Web and serendipitous pure-parallel pointings). However, the more recent results at $z \simeq 13$ from the full 0.54\,deg$^2$ COSMOS-Web area reported by \citet{Franco2025} lie considerably higher, in fact close to our previous $z\simeq11$ result.

At first sight this is surprising because, while they probe a nominally slightly lower redshift range of $11.9<z<13.6$, their median redshift is high ($z_{\rm med}=13.04^{+0.28}_{-0.16}$) and entirely consistent with the median redshift of our $12.5<z<14.5$ bin ($z_{\rm med}=13.1$). However, we caution against over-interpreting the comparison with COSMOS-Web, given that the survey (as utilised by \citealt{Franco2025}) only has four NIRCam filters (F115W, F150W, F277W, F444W), and hence the identification of F150W dropout galaxies relies either on detection in the shallower ground-based UltraVISTA K$_S$ imaging (in lieu of F200W), or comes with very large redshift uncertainties (due to the lack of a solid photometric constraint on the location of the inferred Lyman-break). These limitations may be introducing some unhelpful selection effects, as hinted at by the aforementioned unexpectedly high median redshift of the COSMOS-Web $11.9<z<13.6$ sample, and indeed
the COSMOS-Web based studies note that their F150W-dropout results are more prone to contamination. In the future, more robust results at these redshifts can be anticipated from COSMOS-Web samples that incorporate the F200W+F356W (plus deeper F115W) imaging being delivered by the COSMOS-3D programme.

Although admittedly with some overlaps between fields, all of the results from \citet{Adams2024}, \citet{donnan2024} and \citet{whitler2025} are apparently consistent with the \citet{Franco2025} number densities, but the UV LFs determined by these studies do lie at slightly lower median redshifts. Interestingly, \citet{Weibel2025} and \citet{whitler2025} are similar in having a broader redshift window and $z_{\rm med}=12.7$, $z_{\rm med}=12.8$ respectively, but their derived number densities are in minor ($\simeq1\sigma$) tension. Given the wider area probed by \citet{Weibel2025} (which includes JADES) this perhaps suggests that one/both of the (smaller) JADES fields is over-dense in high-redshift candidates. Overall, in the context of existing literature results at more modest redshifts, our new UV LF at $z \simeq 13.5$, as presented in the left-hand panel of Fig.\,\ref{fig: UV_LF_z13p5}, indicates that the galaxy UV LF starts to decline more dramatically at $z\geq13$.

\begin{table*}
\renewcommand{\arraystretch}{1.2}
\centering
\begin{tabular}{ccccccc}
\hline
Redshift & Function & $M^{\star}$ & $\phi^{\star}$ & $\alpha$ & $\beta$ & $\mathrm \log_{10}{(\rho_{\mathrm{UV}}/{\rm erg\,s}^{-1}{\rm Hz}^{-1}{\rm Mpc}^{-3})}$\\
\hline
13.5 & DPL & $-19.46^{+0.57}_{-0.81}$ & ($2.70^{+6.47}_{-2.11}) \times10^{-5}$ & $-2.35$ (fixed) & $-4.16$ (fixed) & $24.41^{+0.15}_{-0.16}$ \\[1.ex]
13.5 & Schechter & $-19.90^{+0.58}_{-0.92}$ & ($1.58^{+3.89}_{-1.31}) \times10^{-5}$ & $-2.44$ (fixed) & - & $24.42^{+0.14}_{-0.16}$ \\[1.ex]
\hline
13.5 & DPL (pure $M^{\star}$)& $-19.58^{+0.07}_{-0.06}$ & $2.04 \times10^{-5}$ (fixed) & $-2.35$ (fixed) & $-4.16$ (fixed) & $24.37^{+0.04}_{-0.05}$ \\[1.ex]
13.5 & DPL (pure $\phi^{\star}$) & $-20.87$ (fixed) & ($1.77^{+0.28}_{-0.27}) \times10^{-6}$ & $-2.35$ (fixed) & $-4.16$ (fixed) & $24.11^{+0.06}_{-0.07}$ \\[1.ex]
\end{tabular}
\caption{The parameters that are determined for our fiducial double power-law fit to the $z=13.5$ UV luminosity function, fixing both the bright and faint-end slopes to their $z=11$ values from \citet{mcleod2024}. We also include the alternative Schechter function fits, fixing the $\alpha$ slope to the previously determined value for $z=11$. In the lower panel, we include alternative fits assuming pure $\phi^{\star}$ evolution and pure $M^{\star}$ evolution for illustration of how the luminosity function declines between $z=11$ and $z=13.5$. Note that the pure luminosity evolution closely matches that of the fiducial double power-law fit. For each of our fitted functions, we include the integrated luminosity density $\mathrm \log_{10}{(\rho_{\mathrm{UV}})}$, integrated to a limit of $M_{1500}=-17$.}
\label{tab:params}
\end{table*}
\renewcommand{\arraystretch}{1.0}

In the right-hand panel of Fig.\,\ref{fig: UV_LF_z13p5} we compare our new $z \simeq 13.5$ galaxy UV LF with the predictions of several recent theoretical models of galaxy evolution. Interestingly, it can be seen that our $z \simeq 13.5$ LF agrees very well with the predictions of the simple, constant star-formation efficiency, evolving halo mass function model presented by \citet{Donnan2025a}. Rather than requiring any redshift evolution in star-formation efficiency, dust attenuation, or the IMF (modifications which have been invoked by various authors to reproduce the observed high-redshift UV LF and GSMF), this simple model relies on ever-decreasing {\it average} stellar-population ages at extreme redshifts, converging on an epoch of galaxy formation around $z\simeq15$.

The observations also agree well with the attenuation-free model (AFM) predictions at $z=13$ from \citet{Ferrara2023,ferrara2025}, which suggests that a lack of dust at very high redshifts ($z>10$) may play a role in explaining the abundance of UV-bright galaxies. However, we note that significant dust attenuation is not actually expected in the relatively low-mass galaxies that populate the currently observed extreme-redshift UV LF.

By contrast, the \citet{Gelli2024} mass-dependent stochasticity model predictions lie somewhat below the observational data, apparently consistent within $\simeq 1\sigma$ with each bin, but in fact significantly  under-predicting the cumulative galaxy counts at these redshifts. This confirms that increasing burstiness cannot (on its own) explain the high-redshift evolution of the overall galaxy UV LF (albeit it can help to boost the bright end due to increased scatter).

The fiducial model from \citet{mauerhofer2025} at $z=13.89$ more severely under-predicts the number densities that we observe (albeit this could be partly due to the fact the predictions are provided at a slightly higher redshift, and the galaxy evolution could be steepening rapidly). However, this model, when adapted to include an evolving star-formation efficiency, then significantly overshoots the data. From close inspection of Fig.\,\ref{fig: UV_LF_z13p5}, it can be seen that the (apparently) best version of this model is the one that incorporates an evolving IMF. However, somewhat confusingly, it can also be seen that the predictions of the evolving-IMF version of the \citet{mauerhofer2025} model are very similar to those of the \textsc{Astraeus X} model \citep{Hutter2025} at $z=13$ which assumes a non-evolving \citep{salpeter1955} IMF, while the variant of that model which includes a top-heavy IMF this time over-predicts our observations. This discrepancy widens at ever brighter luminosities, suggesting that evolution in the IMF is not required to reproduce the UV LF at these epochs, and indeed can result in poorer predictions.

\begin{figure}
    \centering
    \includegraphics[width=0.5\textwidth]{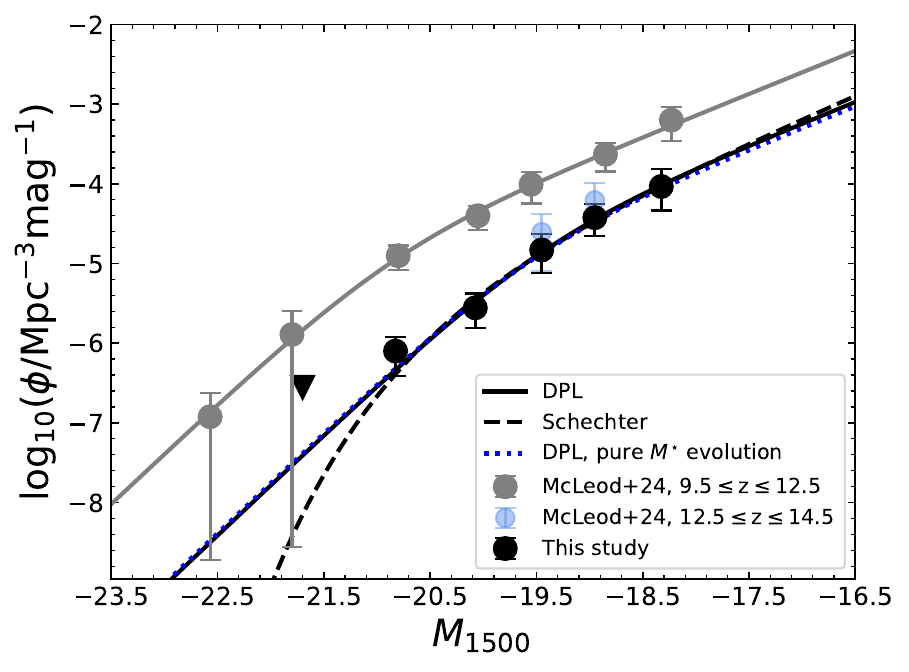}
    \caption{The UV LF at $z=13.5$ with three alternative functional forms over-plotted. With the present data, the Schechter and double power-law fits are indistinguishable. Based on the limits at $M_{1500}=-21.7$, we require another $\gtrsim1$\,dex of survey volume to make progress.}
    \label{fig: LF_different_functions}
\end{figure}

Overall, while our results cannot rule out increased star-formation efficiency at higher redshifts (as has been previously suggested may be necessary to reproduce the observed high-redshift UV LF: e.g., \citealt{Harikane2023,Harikane2025}), we see no evidence that this is required to reproduce the data. Meanwhile, the attenuation-free model, while arguably supported by the observed blue colours of the galaxies in these extreme redshift samples (e.g., \citealt{cullen2023,cullen2024,topping2024}), is also not actually required to explain the prevalence and properties of the majority of galaxies which populate the observed UV LF at $z \geq 11$. Going further, the necessity or otherwise of an evolving IMF also remains unclear, given the conflicting predictions of the \citet{mauerhofer2025} and \citet{Hutter2025} models, as does the requirement to introduce increasingly stochastic star-formation histories
\citep{Gelli2024}. For now, the simple model introduced by \citet{Donnan2025a}, with no modifications in stellar astrophysics, appears as effective as any at explaining the data. Ultimately, the degeneracies between these different theoretical explanations of the data will only be broken by improvements in our knowledge of stellar population ages and the GSMF at very early times.

Finally, we note that the lack of a bright-end anchor in our current $z=13.5$ LF determination prevents us from identifying the functional form with the present data. The upper limit at $M_{1500}=-21.7$ is unable to distinguish between a Schechter function and a double power law. Hence, we quote our fits and integrations to both of these forms in Table \ref{tab:params}, and show both of these functional forms in Fig \ref{fig: LF_different_functions}. As can be seen, it would require a far greater cosmic volume -- corresponding to an additional $\simeq$\,1 dex -- in order to distinguish between these alternative functional forms. However, it can be seen that the faint end of our $z = 13.5$ LF is insensitive to the chosen functional form, and this is reflected in a modest variation in the calculated $\rho_{\rm UV}$ (see Table \ref{tab:params}). We also illustrate that pure $M^{\star}$ evolution, dimming $\simeq1.3$ mag with respect to our $z=11$ determination in \citet{mcleod2024}, is also an acceptable fit to the present data.

\begin{figure*}
    \begin{tabular}{cc}
    \centering
    \includegraphics[width=0.5\textwidth]{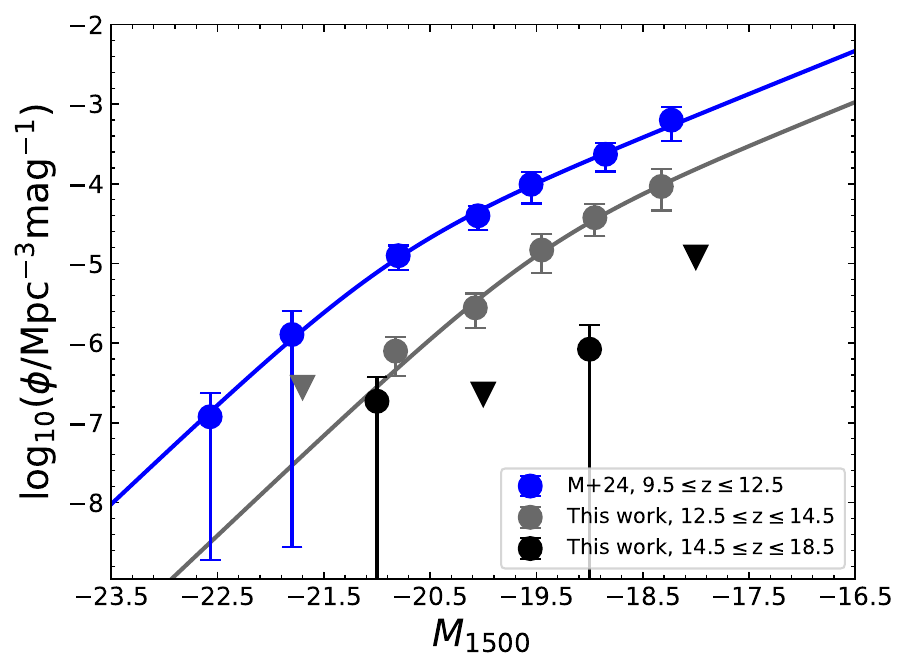} &
    \includegraphics[width=0.5\textwidth]{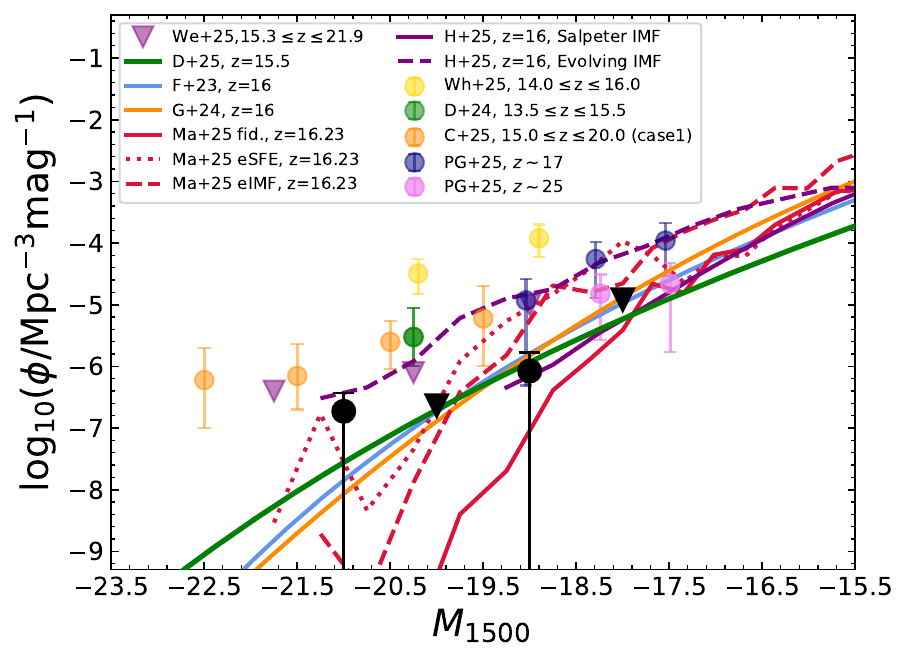}
    \end{tabular}
    \caption{Our determination of the galaxy UV luminosity function at $14.5<z<18.5$. In the left-hand panel, we include a comparison with both the $z\simeq11$ UV LF determination from \citet{mcleod2024}, and the new $z=13.5$ results from the current study. We see a factor $>100\times$ decrease in the number density of galaxies between $z\simeq11$ and our new $z\simeq15.5$ bin. In the right-hand panel, we include a compendium of literature data exploring the $z>13$ Universe. Our derived galaxy number densities are significantly lower than those reported in these other recent studies, with the exception of \citet{Weibel2025}.
    The full references for the literature results summarized in the right-hand panel legend are: Wh+25=\citet{whitler2025}; PG+25=\citet{perezgonzalez2025}; We+25=\citet{Weibel2025}; D+24=\citet{donnan2024}; C+24=\citet{castellano2025}, M+25=\citet{mauerhofer2025}; G+24=\citet{Gelli2024}; F+23=\citet{Ferrara2023}.}
    \label{fig: UV_LF_z15p8}
\end{figure*}

\subsection{Extension to $\mathbf{z\simeq16}$}
While our final sample only contains two $z>14.5$ sources, it is instructive to attempt to place some preliminary constraints on the evolving UV LF at yet earlier cosmic times. We therefore construct the UV LF at $14.5\leq z \leq 18.5$ ($z_{\mathrm{mean}}=15.8$), adopting an identical methodology as for the $z\simeq13.5$ LF. We stress here that the number densities quoted for the $z=15.8$ LF are tentative, with two highly uncertain single-object bins, and should thus be treated as providing upper limits at $z>14.5$, given the lack of any candidates at $z\geq16$ or earlier.

We present our tentative $z=15.8$ UV LF determination in the left panel of Fig. \ref{fig: UV_LF_z15p8}, with our $z=13.5$ UV LF and the \citet{mcleod2024} $z=11$ measurements over-plotted for comparison. As would be expected from the dearth of $z>14.5$ candidates, we find evidence for a continued (indeed, accelerating) rapid decline in galaxy number density with increasing redshift. This can be quantified in terms of either a $\simeq1.6$ mag dimming of $M^{\star}$ (assuming pure luminosity evolution) from $z=13.5$, or a factor $\simeq38\times$ decline in $\phi^{\star}$ (assuming pure density evolution) over the relatively small cosmic time interval of $\simeq100$\,Myr. Both of these interpretations are well-matched to the number density at $M_{1500}=-19$, but neither the brighter bin at $M_{1500}=-21$ nor the upper limits can distinguish between these alternative parameterisations at present. However, as luminosity evolution appears to describe the data well over the redshift range $9.5<z<14.5$, we adopt continued luminosity evolution (for continuity) for the purposes of calculating the inferred star-formation rate density (see Section \ref{section6}).

The limit at $M_{1500}=-20$ suggests that PAN-z16-1, the sole object in the $M_{1500}=-21$ bin, is likely to be anomalously bright for the redshift and survey volume probed, with only one source in the $12.5<z<14.5$ LF being brighter. The alternative possibility is that PAN-z16-1 is a low-redshift interloper (e.g., it lies at the \textsc{LePhare} secondary redshift solution, $z_{\rm sec}\simeq3$), or, as with PAN-z14, it is indeed a high-redshift galaxy, but has had its photometric redshift scattered significantly upwards. The addition of some deep medium-band F182M and/or F210M imaging (to more precisely constrain the Lyman break, and the potentially important effect of damping by the IGM), or deep spectroscopic follow-up with NIRSpec would clarify this.

Our other, fainter $z>14.5$ source, PAN-0332-2754-2030538, which is solely occupying the $M_{1500}=-19$ bin, was found in an ultradeep PANORAMIC pointing. This object has already been reported as a potential $z\simeq16$ candidate by \citet{hainline2026} in GOODS South (their ID 511524). We discuss other $z>14.5$ candidates in the literature in Section \ref{section: comparison_candidates}.

Recently, several studies have claimed the discovery of galaxy candidates at $z>15$, in some cases in surprising numbers at unexpectedly  high redshifts (and hence with surprisingly extreme inferred luminosities; see, e.g., Section \ref{section: comparison_candidates}). We compare our $z \simeq 15$ number density constraints to several of these reported results in the right-hand panel of Fig. \ref{fig: UV_LF_z15p8}. Our new results lie significantly below those reported in these other recent studies (in some cases more than 1 dex lower), with the exception of \citet{Weibel2025} who also find a lack of $z>15$ sources. Although we are cautious not to over-interpret the results (given the large uncertainties on the single-object bins) our number densities and upper limits are in agreement with the predictions of several theoretical models (e.g., \citealt{Ferrara2023,Gelli2024,Hutter2025,Donnan2025a}). Once again, there is little to choose between the models, which all have differing prescriptions for explaining the abundance of the extreme redshift galaxies (e.g., bursty star-formation histories, an evolving IMF,  zero dust attenuation, etc.). There thus remain many degeneracies, but as is the case at  $z=13.5$, the agreement with the simple evolving HMF model from \citet{Donnan2025a} with our results at $z = 15.5$ continues to suggest that current observations can be explained without the introduction of new (or rapidly evolving) astrophysics.

\subsection{Discussion of notable $\mathbf{z\geq15}$ candidates in the literature}
\label{section: comparison_candidates}
Given the inevitable interest in pushing the redshift frontier to "cosmic dawn", numerous teams have sought to uncover galaxy candidates at  $z\geq15$, and claims have now been advanced for many such objects, extending out to redshifts as high as $z \simeq 30$ \citep{gandolfi2026}. Since our own high-redshift galaxy sample, presented here, contains a notable lack of $z>14.5$ galaxies (indeed only two possible candidates remain at $z \simeq 15$), it is important to investigate the potential reasons for such divergent results, especially since several of the more extreme redshift candidates have been purported to lie within some of the survey fields also investigated here. In this subsection we therefore carefully revisit the extreme redshift candidates which have been reported in the recent literature.

First, we consider the surprisingly large number of $z>14$ candidates reported in the recent \citet{hainline2026} JADES DR5 sample. For GOODS-South, our sample includes 3/10 of the reported JADES sources (60936, 506981, 511524). One object, 385830, is in a region for which we had no data, while another object, 184202, was removed due to a marginal F090W detection. The reason we do not recover the remaining five sources appears to be our more conservative SNR$\geq8$ cut and/or objects being excluded in our selection due to plausible lower-redshift solutions. In GOODS-North, we selected 2/7 objects in common with \citet{hainline2026}. Their object 1023895 is known previously from \citet{whitler2025}, while 1204402 is in our sample at a slightly lower $z=12.4$, which appears to be consistent with their secondary p(z) peak. We also recover a possible high redshift for their candidate 1128351, but this source was excluded from our sample as there is only a marginal statistical preference for a high-redshift solution. JADES candidate 1236036 is in a region not covered by our analysis, and the remaining three objects were removed by our conservative SNR$\geq8$ cut.

As one of the deepest NIRCam datasets obtained to date, the GLIMPSE survey, covering the lensing cluster Abell S1063, might naturally be expected to provide excellent prospects for uncovering the earliest galaxies, and indeed \citet{kokorev2025} have reported two 
$z>15$ galaxy candidates in this field. Although these are too faint to be included in our sample (even with 0.2$^{\prime\prime}$-diameter aperture photometry they are $\simeq3-4\sigma$ detections in F277W in our reduction) we nonetheless attempted to verify their high-redshift solutions, to determine whether we would have selected them with less stringent detection thresholds. We find an essentially flat $\chi^2(z)$ distribution for 70467, and a slightly preferred $z_{\rm phot}\simeq15$ solution for 72839, with $\Delta\chi^{2}\simeq3$, and so cannot exclude the low-redshift solution for either object. We note that \citet{kokorev2025} in fact report similarly low values of $\Delta \chi^{2}=1.0$ and $\Delta \chi^{2}=1.2$ between the competing SED solutions for these objects, meaning that neither of these high-redshift candidates can be regarded as robust.

Another {\it JWST} lensing field survey in which a $z\simeq15$ galaxy candidate has been reported is CANUCS, in a parallel field near MACS 0416 \citep{asada2026}. For this source we do find a $z_{\rm phot}=15.0^{+0.6}_{-0.5}$ solution; however, even when incorporating the Technicolor (PI Muzzin) medium-band photometry, the competing secondary solution at $z_{\rm sec}=3.3$ is only very slightly disfavoured ($\Delta\chi^{2}=1.4$). Given how red the galaxy is ($\beta=-1.8$; \citealt{asada2026}), this is perhaps unsurprising, confirming that this galaxy is almost certainly a lower-redshift contaminant.

A further ultradeep NIRCam field within which $z>15$ candidates have been reported is NGDEEP, in a region originally known as the HUDFpar2 field in the pre-\textit{JWST} era. The object labelled NGD-z15a was a $z\simeq15$ candidate first reported in \citet{austin2023} and later in \citet{conselice2024}. This object was also selected in \citet{cullen2024} as an apparently robust high-redshift candidate, NGDEEP-1003576. However, for all of these studies, only the first epoch of NIRCam imaging was available. With the addition of NGDEEP epoch 2 and MIDIS imaging, pushing the imaging $\simeq0.6$ mag deeper, the F150W imaging revealed a clear detection, pushing the photometric redshift solution firmly towards that of a dusty $z\simeq3-4$ interloper.

Finally, combining NGDEEP and MIDIS, \citet{perezgonzalez2025} identified a sample of $z=17-25$ galaxies. The F277W-dropouts claimed by \citet{perezgonzalez2025} cannot be selected in our current study by design, as these will naturally fail any $>8\sigma$ criterion in F200W and F277W. However, even using 0.2$^{\prime\prime}$-diameter apertures, we cannot reproduce the high SNR values reported in \citet{perezgonzalez2025} for any of their candidates, suggesting potential differences in our reduction and/or the methods used to measure photometric uncertainties. Nonetheless, we find that z25-1 is clearly detected ($2.9\sigma$) in F277W, in clear tension with their claimed $z\simeq25$ solution. Their sources z25-2 and z25-3 fail to be detected at $>3\sigma$ in F356W. Likewise, z17-1, z17-2 and z17-5 are only detected at $\simeq3\sigma$ in F277W using our reductions.

\begin{figure}
    \centering
    \includegraphics[width=0.5\textwidth]{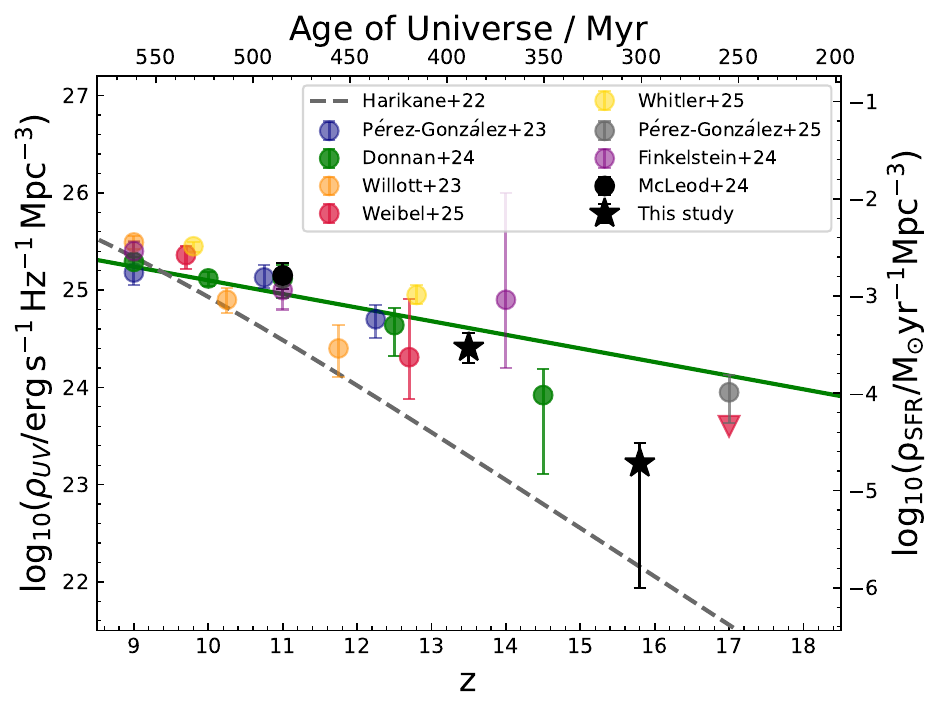}
    \caption{The evolution of the cosmic star-formation rate density with redshift, as measured by this study in the context of other observations at $z>9$. We include the smoothly declining, log-linear $\mathrm{\log_{10}\rho_{\rm UV}(z)}$ relationship from \citet{donnan2024} which closely matches existing observations up to $z\simeq12$, as well as the rapidly declining model from \citet{Harikane2022}. Our $z=13.5$ results lie slightly below the smooth log-linear relation, but remain broadly consistent with earlier studies at $z<13$. However, our lack of $z>14.5$ candidates results in an accelerated decline in $\mathrm{\log_{10}\rho_{\rm UV}(z)}$ at $z>13$, similar to that seen by \citet{Weibel2025}.}
    \label{fig: SFRD}
\end{figure}

\begin{figure*}
    \begin{tabular}{cc}
    \centering
    \includegraphics[width=0.5\textwidth]{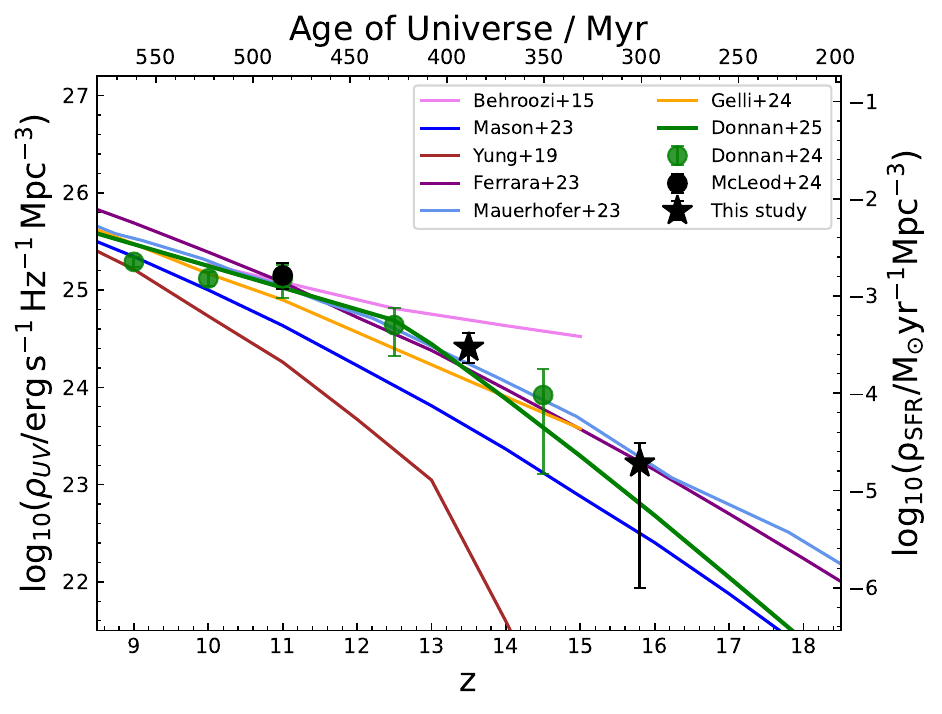} &
    \includegraphics[width=0.5\textwidth]{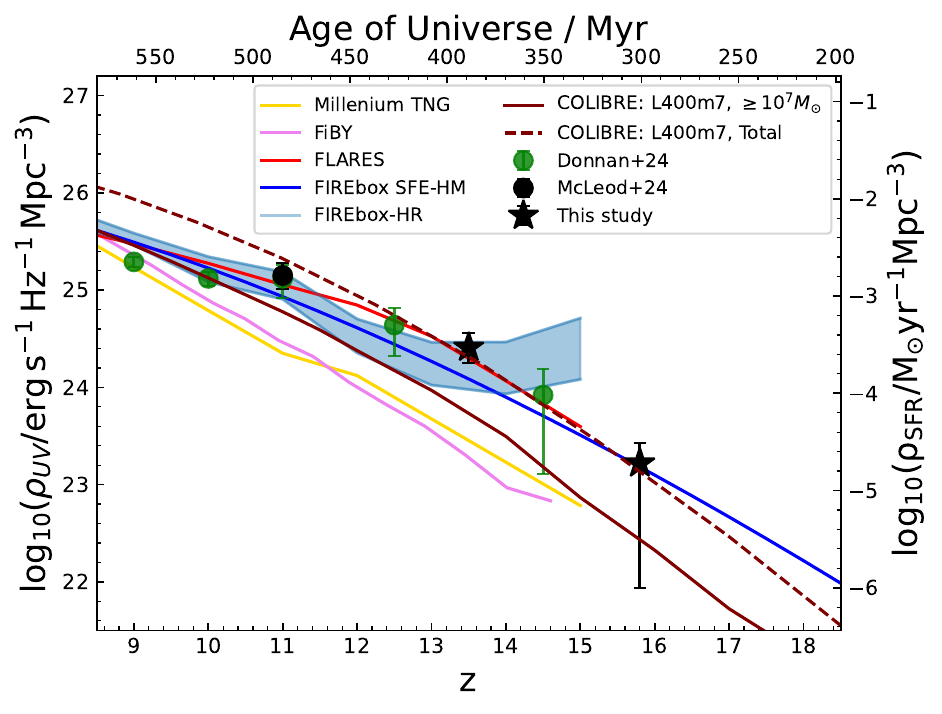}
    \end{tabular}
    \caption{The evolving cosmic star-formation rate density as measured in this study and previously by \citet{donnan2024} in the context of theoretical models (left-hand panel) and simulations (right-hand panel). Our measurements lie above some of the models (e.g., \citet{Yung2019,Mason2023} and simulations such as Millenium TNG \citep{Kannan2023} and FiBY (e.g., \citealt{johnson2013,Paardekooper2013,paardekooper2015}. However, our results are in agreement with the evolving halo mass function model from \citet{Donnan2025a}, the AFM model from \citet{Ferrara2023}, DELPHI \citep{mauerhofer2023} and the FLARES \citep{Wilkins2023} simulation. The mass-dependent stochasticity model from \citet{Gelli2024} lies slightly below our measurements but is still consistent with the data given the uncertainties. The COLIBRE \citep{schaye2025,chaikin2025} simulation closely matches our present results when considering the total possible SFRD. However, when restricting to the $\geq10^{7}M_{\odot}$ population, which would correspond more closely to an integral limit of $M_{\rm UV}=-17$, the results are underestimated. Finally, the FIREbox \citep{feldman2025} high-resolution simulation results are consistent with our measurements in the overlapping redshift range: the theoretical model linking the star-formation efficiency versus halo mass relation from the simulation lies slightly below our $z=13.5$ measurement, but matches well at $z\simeq16$.}
    \label{fig: SFRD_models}
    \end{figure*}

\begin{figure}
    \centering
    \includegraphics[width=0.5\textwidth]{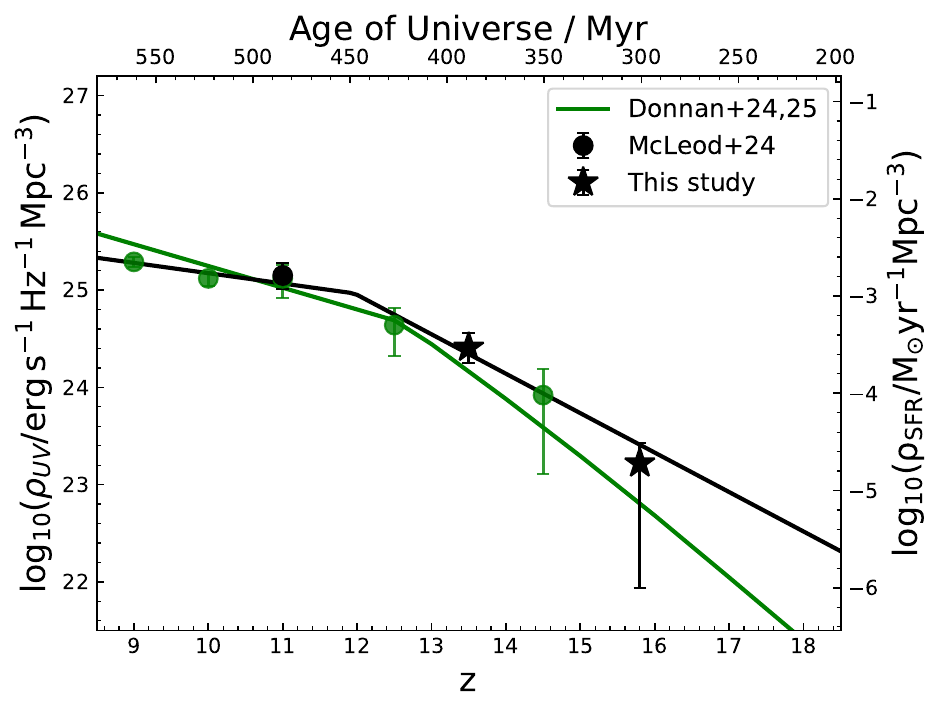}
    \caption{This plot demonstrates that the evolving halo mass function model with progressively younger stellar ages from \citet{Donnan2025a} is well-matched to the observations from \citet{mcleod2024} and \citet{donnan2024} as well as with the new measurements from the current study. Similarly, a simple piecewise power-law, fitted to the data points, provides a good description of the trend in $\mathrm{\log_{10}\rho_{\rm UV}(z)}$ at $z\gtrsim9$, with a break redshift around $z=12$ beyond which the gradient steepens by a factor $\simeq4$ (from $-0.11z$ to $-0.41z$).}
    \label{fig: SFRD_z15p5_HMFmodel}
\end{figure}

\section{Cosmic star-formation rate density}
\label{section6}
Finally, we proceed to calculate the evolving UV luminosity density, $\rho_{\rm UV}$, by performing a luminosity-weighted integral of our derived UV LFs down to a luminosity limit corresponding to $M_{1500}=-17.0$. We include the resulting $\rho_{\rm UV}$ values at $z = 13.5$ in Table \ref{tab:params} for our fiducial double power-law fit, as well as the alternative fit assuming a Schechter function. For completeness, we also provide $\rho_{\rm UV}$ for the pure $M^{\star}$ and $\phi^{\star}$ evolution scenarios. As would be expected, given the constraints provided by the data, these are all very similar, with the exception of the (disfavoured) pure $\phi^{\star}$ evolution, which is lower by a factor of $\simeq 2$. To convert our UV luminosity densities to star formation rate densities, $\rho_{\rm SFR}$, we use the conversion factor $\mathrm{\mathcal{K}_{UV}=1.15\times10^{-28} M_{\odot}\,yr^{-1}/erg/s^{-1}Hz^{-1}}$ as defined in \citet{Madau2014}, which is based on a \citet{salpeter1955} IMF.

Our derived, evolving $\rho_{\rm SFR}$ can be viewed in the context of other recent observational results in Fig.\,\ref{fig: SFRD}. Our measurement at $z\simeq13.5$ is consistent with that derived by \citet{donnan2024} and with the results recently obtained by several other groups at similar redshifts (e.g., \citealt{finkelstein2024,Weibel2025}). Although we do not plot it for clarity, the present results are also consistent with the previous tentative $z=13.5$ measurement of $\rho_{\rm UV}=24.54^{+0.16}_{-0.25}$\,erg\,s$^{-1}$\,Hz$^{-1}$\,Mpc$^{-3}$ published by \citet{mcleod2024}. There is thus, now, an emerging consensus over the value of UV-luminous $\rho_{\rm SFR}$ at $z \simeq 12-14$, with our results  and those of \citet{donnan2024} and \citet{Weibel2025} suggesting a more accelerated decline to earlier times $z\geq13$ than predicted by extrapolation of the smooth log-linear relation fitted by \citet{donnan2024} over the redshift range $z=9-12$.

Although our UV LF at earlier times ($14.5<z<18.5$) is at present inevitably more uncertain, we find that the large drop in the number density of higher redshift candidates translates to a sharp decline in $\rho_{\rm UV}$ out to $z\simeq15.8$, with $\rho_{\rm UV}=23.22^{+0.21}_{-1.28}$\,erg\,s$^{-1}$\,Hz$^{-1}$\,Mpc$^{-3}$ assuming pure luminosity evolution (which we adopt here given the excellent agreement with pure luminosity evolution seen over $9.5<z<14.5$). Our results are similar to the upper limit at $z\simeq17$ provided by \citet{Weibel2025}. Overall, our results reinforce the increasing departure from the well-established smooth decline in star-formation rate density that has been seen out to $z\simeq12$ in early \textit{JWST} studies (e.g., \citealt{Donnan2023a,Finkelstein2023,finkelstein2024,Perez-Gonzalez2023,mcleod2024}), indicative of an increasingly dramatic change at $z > 14$. 
We note that assuming pure $\phi^{\star}$ evolution would result in an even sharper descent at these extreme redshifts, yielding  $\rho_{\rm UV}=22.8^{+0.3}_{-1.1}$\,erg\,s$^{-1}$\,Hz$^{-1}$\,Mpc$^{-3}$ at $z \simeq 15.8$.

In Fig.\,\ref{fig: SFRD_models} we place our $\rho_{\rm UV}$ measurements in the context of various theoretical models (left-hand panel) and simulations (right-hand panel). The measurements at $z>12$ continue to lie above the more rapid decline in $\rho_{\rm SFR}$ predicted from the pre-\textit{JWST} work by \citet{Yung2019}, the Millenium TNG \citep{Kannan2023} and FiBY \citep{paardekooper2015} simulations, and the early-\textit{JWST} forecasts from \citet{Mason2023}. As anticipated from the agreement in the UV LFs, we find that our observations align with the evolving $\rho_{\rm UV} (\rm z)$ from the \citet{Ferrara2023} and \citet{Donnan2025a} models, as well as with DELPHI \citep{mauerhofer2023}. Although our $z=13.5$ data lie slightly above the predictions from \citet{Gelli2024}, our observations are still broadly consistent with their predicted trend. Extrapolating their model to $z\simeq16$, the predictions would be expected to agree with our $z=15.8$ observations. Although the COLIBRE L400m7 \citep{schaye2025,chaikin2025} total possible $\rho_{\rm SFR}$ agree with our observations, these under-predict the observations when including a cut-off stellar mass $\geq10^{7}{\rm M_{\odot}}$, which would be expected to correspond roughly to the $M_{\rm 1500}<-17$ integration limit adopted in our analysis. Over the overlapping redshift range out to $z=13.5$, both FLARES \citep{Wilkins2023} and the high-resolution FIREbox \citep{feldman2025} simulations are consistent with our observations. Although lying below our $z=13.5$ measurements, the FIREbox theoretical model linking star-formation efficiency and halo mass from the simulation is consistent with our $z=15.8$ observations.

Our new observational results are thus consistent with a number of theoretical models of galaxy evolution which have incorporated a range of treatments/modifications in an attempt to explain the prevalence of UV-bright galaxies at least out to $z \simeq 12$
(e.g., increased star-formation efficiency, stochastic star-formation histories, an evolving stellar
initial mass function, and/or a shift towards attenuation-free stellar populations). However, our new 
results, in tandem with our previous measurements from \citet{mcleod2024} and \citet{donnan2024}, are also entirely consistent with a relatively simple galaxy evolution model with no such adjustments, in which the rapid evolution of the dark matter halo mass function at early times is temporarily partially masked by progressively younger stellar population ages, with the inferred
epoch of first galaxy formation lying at $z\simeq15$ \citep{Donnan2025a}. We show how well this simple model describes the data from $z \simeq 9$ to $z \simeq 16$ in Fig.\,\ref{fig: SFRD_z15p5_HMFmodel}.

In Fig.\,\ref{fig: SFRD_z15p5_HMFmodel}, we also plot a piecewise log-linear fit to the measured evolution of $\rho_{\rm UV}(z)$ at $z>9$, fitting our new measurements as well as those derived by \citet{mcleod2024} and \citet{donnan2024}. We find:
\begin{equation*}
\begin{aligned}
\rho_{\rm UV}=-0.11(z - z_{b})+\rho_{UV,b} && z<\,z_{b} \\
\rho_{\rm UV}=-0.41(z - z_{b})+\rho_{UV,b} && z\gtrsim\,z_{b}
\end{aligned}
\end{equation*}

\noindent with the break in the relation at $z_{b}=11.96$ and $\rho_{UV,b}=24.97$. At $z>12$, we find that the decline in $\rho_{\rm UV}(z)$ is around a factor of four steeper than at later cosmic times.

Given the fitted break redshift lies at $z=12$, it is tempting to suggest that this could signal an important transition point in cosmic history, whereby we are exiting a period of very rapid star formation and stellar mass assembly and moving into a phase of more gradual growth/evolution at "lower" redshifts. However, we note that the redshift windows for UV LF determination probed by \textit{JWST}'s typical NIRCam broadband filter set (occasionally including F410M) are broad. For example, in \citet{mcleod2024} the redshift window for the $z=11$ UV LF was $9.5<z<12.5$, in \citet{donnan2024} the $z=12.5$ bin spanned $11.5<z<13.5$, and in the present study, our main constraint is over the redshift range $12.5<z<14.5$. Hence, it would be instructive for future studies to better refine our measurements of the UV LF and $\rho_{\rm UV}$ at $z=12$.

The prospects for doing this with upcoming data sets are exciting. The Cycle 4 PID 7814 program MINERVA \citep{muzzin2025} is currently delivering NIRCam medium-band observations over four of the CANDELS \citep{Grogin2011} fields: UDS, COSMOS, EGS and GOODS North, providing an excellent complement to PRIMER, JADES and CEERS and enabling more precise selection of high-redshift candidates. In particular, EGS will also benefit from further medium-band coverage from SPAM (PID 8559; PIs Davis, Larson). Moreover, given the lack of robust, relatively bright $z>14.5$ candidates uncovered in this study, with only a couple of potential prospects for more speculative (and expensive) spectroscopic followup, it is perhaps prudent for future work to prioritise the consolidation of our understanding of the $z=12-14$ galaxy population, particularly as there still remain many outstanding questions regarding the properties of these early galaxies, such as their dust content, ages and star-formation histories, metallicities and contribution to the early progress of cosmic reionization.

\section{conclusions}
\label{section7}
In this study, we have presented an extensive search for $z\geq12$ galaxy candidates over $>0.6$ sq. degrees and $>150$ sightlines of \textit{JWST} NIRCam imaging. From this search, we have constructed a robust sample of 55 $z\geq11.5$ candidates selected at $\geq8\sigma$, including 33 candidates at $z\geq12.5$. Notably, one of our candidates, PAN-z14-1, has recently been confirmed with NIRSpec to lie at $z_{\rm spec}=13.5$ \citep{donnan2026}, and we await spectroscopic confirmation of a moderately-lensed ($\mu=3.2$) $z\simeq13$ source PLCKG165-2047505, for which we have been awarded observation time with NIRSpec in Cycle 5 (PID 11171; PIs McLeod, Garuda). 

With this sample, we have determined the galaxy UV luminosity function over $12.5<z<14.5$, spanning a dynamic range of $\simeq3$ magnitudes in $M_{\rm 1500}$. Despite our wide-area search, due to the lack of bright galaxies at $M^{\star}<-21$ we are unable to distinguish between a double power-law and Schechter functional form with the present data. For continuity with early \textit{JWST} studies, we adopt the double power-law fit as fiducial. We find that the number densities of galaxies at $z>12.5$ have decreased by a factor of $\simeq12$ with respect to our previous UV LF determination over $9.5<z<12.5$ \citep{mcleod2024}. Alternatively, assuming pure luminosity evolution in the double power-law fit to the data, we find that the evolution can be described by a dimming of $\simeq1.3$ magnitudes in $M^{\star}$. With only $\simeq100\,\rm Myr$ between $z=11$ and $z=13.5$, this demonstrates a more accelerated decline in star formation activity as one probes to earlier cosmic times beyond $z>12$.

We find a dearth of $z>14.5$ galaxy candidates, with only two reported around $z\simeq15.8$ from the PANORAMIC survey area. Our tentative measure of the UV LF, comprised of single-object bins and upper limits, seems to suggest an even more rapid descent in the UV LF. Our results lie below other works in the literature reporting $z>15$ candidates over more modest survey areas.

By integrating our UV LFs, we have placed constraints on the UV luminosity density and hence star formation rate density at $z>12$. We find a rapid decline in the $\rm \rho_{SFR}(z)$ relation at $z>12$, departing from the log-linear relation reported in earlier \textit{JWST} works \citep{donnan2024}. We find $\rm \rho_{SFR}(z)$ is better represented by a piecewise log-linear fit, with a break redshift of $z\simeq12$, where the slope steepens by a factor of $\simeq4$.

Our results over $12.5<z<18.5$ are consistent with a number of theoretical models offering different treatments for, e.g., evolving IMF, mass-dependent stochasticity and dust attenuation-free stellar populations at higher redshift. While many of these models offering differing prescriptions to reproduce the abundance of UV-bright $z>10$ galaxies are consistent with the data, the simple model by \citet{donnan2024} where one adopts a rapidly evolving dark matter halo mass function out to earlier redshifts, partially masked by progressively younger stellar ages up to a typical formation epoch at $z\simeq15$ without the need of altering star formation and stellar population properties, is well-matched to the present observations.

\section*{data availability}
At the time of writing, the data sets used in this manuscript were all publicly available from the \textit{JWST} archive on MAST. For JADES and CEERS, these are available as part of their respective public data releases.

\section*{acknowledgements}

We would like to express our appreciation to all who have worked to enable the many \textit{JWST} NIRCam surveys utilised in this work. We also warmly thank Andrea Ferrara, Viola Gelli, Robert Feldman, Anne Hutter, Joop Schaye, Evgenii Chaikin and Pratika Dayal for providing theoretical model/simulation predictions, Andrea Weibel for providing observational results, Adi Zitrin for providing a MACS 2135 lensing map, and Jindra Gensior for useful discussions.

DJM and JSD acknowledge the support of the Royal Society through the award of a Royal Society University Research Professorship to JSD.
RJM acknowledges the support of the UK Science and Technology Facilities Council (STFC). ACC acknowledges support from a UKRI Frontier Research Grantee Grant (PI Carnall; grant reference EP/Y037065/1). FC and TMS acknowledge support from a UKRI Frontier Research Guarantee Grant (PI Cullen; grant reference: EP/X021025/1).

This work is based primarily on observations made with the NASA/ESA/CSA {\it James Webb Space Telescope}. The data were obtained from the Mikulski Archive for Space Telescopes at the Space Telescope Science Institute, which is operated by the Association of Universities for Research in Astronomy, Inc., under NASA contract NAS 5-03127 for {\it JWST}. These observations used in this study are associated with programs 1063, 1176, 1180, 1181, 1199, 1208, 1210, 1283, 1286, 1324, 1345, 1355, 1433, 1727, 1810, 1837, 1840, 1895, 1963, 1967, 2079, 2234, 2282, 2514, 2555, 2561, 2566, 2727, 2732, 2736, 2738, 2739, 2750, 2756, 2767, 2883, 3073, 3215, 3293, 3362, 3516, 3538, 3859, 3990, 4043, 4111, 4125, 4212, 4287, 4598, 4625, 4744, 4446, 5398, 5893, 6368, 6434, 6511, 6541, 6549, 6556, 6564, 6882. The authors acknowledge the associated teams for developing their observing programs. In particular, we appreciate the programs that were developed and executed with a zero-exclusive-access period.

This work is based (in part) on observations taken by the RELICS Treasury Program (GO 14096) with the NASA/ESA {\it Hubble Space Telescope}, which is operated by the Association of Universities for Research in Astronomy, Inc., under NASA contract NAS5-26555.

Some of the lensing clusters analysed come from the CLASH program with the NASA/ESA {\it Hubble Space Telescope} (GO 12065). The mass models were constructed by \citet{zitrin2009,zitrin2013,Zitrin2015}, and obtained through the Hubble Space Telescope Archive, as a high-end science product of the CLASH program \citep{Postman2012}.

This work utilizes gravitational lensing models produced by PIs Bradač, Natarajan \& Kneib (CATS), Merten \& Zitrin, Sharon, Williams, Keeton, Bernstein and Diego, and the GLAFIC group. This lens modelling was partially funded by the HST Frontier Fields program conducted by STScI. STScI is operated by the Association of Universities for Research in Astronomy, Inc. under NASA contract NAS 5-26555. The lens models were obtained from the Mikulski Archive for Space Telescopes (MAST).

This research has made use of the SVO Filter Profile Service (http://svo2.cab.inta-csic.es/theory/fps/) supported from the Spanish MINECO through grant AYA2017-84089 \citep{Rodrigo2012} \citep{Rodrigo2020}.

This research has made use of NASA's Astrophysics Data System Bibliographic Services.

\bibliographystyle{mnras}
\bibliography{LF_McLeod_2026}

\appendix

\section{DESCRIPTION OF THE DATASET}
In Tables \ref{table: depths},\ref{table: depths2}, and \ref{table: depths3}, we provide the global $5\sigma$ limiting magnitudes through each of the relevant broad-band NIRCam filters (and F410M) for each of the survey fields included in this study.
\begin{table*}
    \centering
        \caption{The 5$\sigma$ global limiting magnitudes in the broadband (and F410M) NIRCam images for each of the fields explored in this study. These are measured in 0.20-arcsec diameter apertures and corrected to total using a point-source correction. Note that some fields have significant spatial variations in depth, due to enhanced background provided by foreground sources, for example in cluster fields.\\
        $^\dag$ indicates the field is not included in UV LF, because of a lack of lensing map.}
    \begin{tabularx}{\textwidth}{l|XXXXXXXXXXXXXX}
        \hline
        Field & F070W & F090W & F115W & F150W & F200W & F277W & F356W & F410M & F444W\\
        \hline
        J1235 & 28.9 & 28.7 & 29.0 & 29.0 & 29.4 & 29.5 & 29.5 & - & 28.8 \\
        CLG 1212 & - & 28.1 & - & 28.3 & 28.5 & 28.4 & 28.6 & - & 28.3 \\
        El Gordo & - & 28.3 & 28.6 & 28.5 & 28.8 & 28.8 & 28.8 & 28.3 & 28.5 \\
        GAMA-100033$^\dag$ & - & 28.2 & - & 28.4 & 28.6 & 28.6 & 28.7 & - & 28.4 \\
        PLCKG 165 & - & 28.4 & 28.5 & 28.7 & 28.9 & 28.9 & 28.9 & 28.3 & 28.6 \\
        PLCKG 191 & - & 28.2 & 28.4 & 28.4 & 28.7 & 28.7 & 28.7 & 28.2 & 28.4 \\
        PEARLS ERS & - & 28.6 & 28.9 & 28.8 & 29.0 & 29.0 & 29.0 & 28.5 & 28.7 \\
        MACS 0416 & - & 28.9 & 29.1 & 29.2 & 29.3 & 29.2 & 29.3 & 28.5 & 28.9 \\
        JADES/GOODS-S & - & 29.4 & 29.8 & 29.9 & 29.9 & 30.1 & 30.1 & 29.5 & 29.4 \\
        JADES/GOODS-N & - & 28.9 & 29.3 & 29.3 & 29.4 & 29.5 & 29.4 & 28.9 & 29.0 \\
        MACS 1149 & - & 28.7 & 28.9 & 29.0 & 29.1 & 29.0 & 29.1 & 28.6 & 28.7 \\
        MACS 0417 & - & 28.6 & 28.8 & 29.0 & 29.1 & 28.9 & 29.0 & 28.6 & 28.7 \\
        MACS 1423 & - & 28.7 & 28.9 & 29.1 & 29.3 & 29.1 & 29.2 & 28.6 & 28.8 \\
        Abell 370 & 28.8 & 28.7 & 28.7 & 28.9 & 28.9 & 29.0 & 29.0 & 28.3 & 28.7 \\
        UNCOVER/Abell 2744 & 29.0 & 29.1 & 29.0 & 29.1 & 29.2 & 29.2 & 29.3 & 28.6 & 28.9 \\
        CEERS/EGS & - & 28.8 & 29.2 & 29.1 & 29.4 & 29.3 & 29.3 & 28.5 & 28.8 \\
        SGAS 1226 & - & - & 25.9 & 26.2 & 26.3 & 26.9 & 27.0 & - & 26.7 \\
        SGAS 1723 & - & - & 25.8 & 26.2 & 26.3 & 27.1 & 27.1 & - & 27.0 \\
        SGAS 1050 & - & - & 27.4 & 27.7 & - & 28.1 & 28.1 & - & 27.8 \\
        SPT 0418$^\dag$ & - & - & 27.4 & 27.0 & 27.5 & 28.3 & 27.8 & - & 27.6 \\
        MACS 0647 & - & - & 28.1 & 28.3 & 28.6 & 28.3 & 28.4 & - & 28.1 \\
        Blue Jay South & - & 28.8 & 29.3 & 29.6 & 29.8 & - & 29.2 & 29.3 & 29.2 \\
        PRIMER/COSMOS & - & 28.1 & 28.0 & 28.2 & 28.7 & 28.6 & 28.8 & 28.1 & 28.2  \\
        PRIMER/UDS & - & 27.9 & 28.1 & 28.4 & 28.5 & 28.6 & 28.6 & 27.9 & 28.1  \\
        Abell 1689$^\dag$ & - & - & 27.4 & 28.1 & 27.2 & 27.7 & 27.6 & - & 27.9 \\
        B14-65666 & - & - & 26.9 & 28.3 & 27.2 & 27.9 & 28.0 & - & 28.2 \\
        BDF3299 & - & - & 28.1 & 28.5 & 28.2 & 28.8 & 28.7 & - & 27.8 \\
        SXDF-NB1006-2 & - & - & 27.1 & 28.6 & 27.4 & 28.2 & 28.1 & - & 28.4 \\
        NGDEEP+MIDIS & - & - & 30.3 & 30.3 & 30.5 & 30.4 & 30.3 & 29.9 & 30.3 \\
        WHL 0137 & - & 28.1 & 28.4 & 28.6 & 28.6 & 28.7 & 28.7 & 28.1 & 28.4 \\
        SDF-LBG-ID34 & - & - & 27.1 & 27.1 & 27.6 & - & - & - & 27.6 \\
        J0235-0532 & - & - & 26.7 & 26.3 & 27.0 & - & - & - & 27.1 \\
        J1211-0118 & - & - & 26.8 & 27.3 & 27.6 & - & - & 27.1 & 27.1 \\
        J1241+2219-B$^\dag$ & - & - & 27.3 & 27.6 & 27.6 & - & - & - & - \\
        J2236+0032 & - & - & 29.1 & 27.4 & 29.2 & - & 28.3 & - & 28.9 \\
        Cartwheel & - & 27.5 & - & 27.9 & 28.1 & 28.1 & 28.2 & - & 27.9 \\
        Stephan's Quintet & - & 27.1 & - & 27.5 & 27.6 & 27.9 & 28.0 & - & 27.7 \\
        SMACS 0723 & - & 28.8 & 29.1 & 29.0 & 29.1 & 29.0 & 29.1 & - & 29.0 \\
        NEP TDF & - & 28.6 & 28.8 & 28.9 & 29.0 & 29.2 & 29.2 & 28.5 & 28.7 \\
        RXJ 2129 & - & - & 27.8 & 28.6 & 28.2 & 28.4 & 28.8 & - & 28.2 \\
        SPT 0615-57 & - & 28.3 & 28.5 & 28.8 & 28.9 & 28.9 & 28.9 & 28.3 & 28.4 \\
        Sunburst arc$^\dag$ & - & - & 27.5 & 27.7 & 27.9 & 28.0 & 28.2 & - & 28.0 \\
        MACS 0138/Requiem$^\dag$ & - & - & 27.9 & 27.8 & 27.9 & 28.2 & 28.3 & - & 28.2 \\
        GLASS/Abell 2744 & - & 29.3 & 29.5 & 29.2 & 29.5  & 29.4 & 29.4 & - & 29.5 \\
        GLASS-par-3073 & - & - & 29.3 & 29.5 & 29.5 & 29.3 & 29.3 & - & 28.0 \\
        NGC 2936 & - & 27.5 & - & 27.9 & 28.1 & 28.3 & 28.3 & - & 28.0 \\
        IC 4553 & - & 27.2 & - & 27.5 & 27.6 & 27.9 & 28.0 & - & 27.8 \\
        Arp 107 & - & 27.4 & - & 27.9 & 28.0 & 28.2 & 28.2 & - & 27.8 \\
        Bullet Cluster$^\dag$ & - & 28.5 & 28.7 & 28.8 & 28.8 & 28.6 & 28.7 & 28.3 & 28.5 \\
        GLIMPSE/Abell S1063 & - & 29.8 & 29.9 & 29.9 & 29.9 & 29.9 & 30.0 & 29.5 & 30.0 \\
        GOODS-South/GO-1286 & 29.1 & 29.4 & 29.6 & 29.7 & 29.7 & 29.8 & 29.7 & 29.3 & 29.5 \\
        Sapphires/MACS 0416 & 29.5 & 29.4 & 29.4 & 29.6 & 29.7 & 29.6 & 29.6 & 29.1 & 29.4 \\
        POPPIES/Abell S1063 & 29.1 & 29.3 & 29.4 & 29.5 & 29.7 & 29.2 & 29.4 & - & 29.4 \\
        POPPIES+Sapphires/EGS & - & - & 29.5 & 29.7 & 29.9 & 29.7 & 29.7 & - & 29.6 \\
        CEERS-East & - & 29.0 & 29.3 & 29.5 & 29.7 & 29.5 & 29.5 & 28.1 & 29.1 \\
        Sapphires 09p59+02p35 & 28.8 & 29.0 & 27.7 & 28.0 & 29.0 & 28.3 & 29.2 & 28.5 & 27.8 \\
        Cosmic Eye/MACS 2135 & - & - & 27.4 & 27.6 & - & 28.0 & 28.1 & - & 27.8 \\
        GO-4625/COSMOS & - & 28.5 & 28.7 & 28.9 & 29.1 & 28.9 & 29.3 & - & 28.6 \\
        \hline
    \end{tabularx}
    \label{table: depths}
\end{table*}

\begin{table*}
    \centering
    \caption{Continued}
    \begin{tabularx}{\textwidth}{l|XXXXXXXXXXXXXX}        \hline
        Field & F070W & F090W & F115W & F150W & F200W & F277W & F356W & F410M & F444W\\
        \hline
        PAN-00:03$+$11:19 & - & -& 27.9 & 28.1 & 28.3 & 28.5 & 28.5 & - & 28.1 \\
        PAN-01:04$-$55:08 & - & - & 27.4 & 27.7 & 27.9 & 28.4 & 28.4 & - & 28.1 \\
        PAN-01:04$+$02:16 & - & - & 29.0 & 28.9 & 29.1 & 29.0 & 29.0 & 28.9 & - \\
        PAN-01:34$-$15:31 & - & - & 27.4 & 27.7 & 27.9 & 28.3 & 28.3 & - & 28.0 \\
        PAN-01:38$-$10:19 & - & - & 28.8 & 29.0 & 29.1 & 29.0 & 29.1 & - & 28.7 \\
        PAN-01:44$+$17:14 & - & -& 28.8 & 29.0 & 29.2 & 29.1 & 29.1 & - & 28.7 \\
        PAN-02:17$-$02:13 & - & - & 29.2 & 29.3 & 28.5 & 28.8 & 29.3 & - & 28.9 \\
        PAN-02:17$-$05:19 & - & - & 28.5 & 28.7 & 28.7 & 28.7 & 28.8 & - & 28.5 \\
        PAN-02:17$-$05:21 & - & - & 28.5 & 28.7 & 28.7 & 28.7 & 28.8 & - & 28.5 \\
        PAN-02:18$-$05:16 & - & - & 28.4 & 28.6 & 28.6 & 28.6 & 28.6 & - & 28.3 \\
        PAN-03:14$-$17:58 & - & - & 28.3 & 28.5 & 28.7 & 28.9 & 28.9 & - & 28.5 \\
        PAN-03:32$-$27:54 & - & - & 30.3 & 29.8 & 29.5 & 29.6 & 29.6 & 29.6 & 29.5 \\
        PAN-03:32$-$27:57 & - & - & 28.6 & 28.8 & 28.8 & 28.8 & 28.9 & - & 28.6 \\
        PAN-04:16$-$24:10 & - & - & 28.5 & 28.7 & 28.9 & 28.9 & 28.9 & - & 28.5 \\
        PAN-04:38$-$68:49 & - & - & 28.0 & 28.2 & 28.4 & 28.6 & 28.6 & - & 28.2 \\
        PAN-09:31$+$08:19 & - & - & 27.8 & 28.1 & 28.3 & 28.4 & 28.5 & - & 28.0 \\
        PAN-09:42$+$09:22 & - & - & 27.9 & 28.1 & 28.3 & 28.5 & 28.5 & - & 28.1 \\
        PAN-10:00$+$02:07 & - & - & 28.4 & 28.6 & 28.6 & 28.6 & 28.7 & - & 28.3 \\
        PAN-10:00$+$02:08 & - & - & 28.4 & 28.6 & 28.6 & 28.6 & 28.8 & - & 28.5 \\
        PAN-10:00$+$02:41 & - & - & 27.3 & 27.5 & 27.7 & 28.1 & 28.1 & - & 27.7 \\
        PAN-10:07$+$21:09 & -  & - & 27.8 & 28.1 & 28.3 & 28.4 & 28.4 & - & 28.0 \\
        PAN-12:19$+$03:29 & - & - & 27.7 & 27.9 & 28.1 & 28.3 & 28.3 & - & 27.9 \\
        PAN-12:31$-$16:18 & - & -& 27.4 & 27.7 & 27.8 & 28.2 & 28.2 & - & 27.8 \\
        PAN-12:37$+$62:10 & - & - & 28.6 & 28.9 & 28.8 & 28.9 & 29.0 & - & 28.6 \\
        PAN-12:38$+$62:13 & - & - & 28.5 & 29.2 & 28.7 & 28.9 & 28.5 & - & 28.6 \\
        PAN-12:56$+$56:51 & - & - & 27.8 & 28.0 & 28.2 & 28.4 & 28.7 & - & 28.3 \\
        PAN-13:00$+$12:14 & - & - & 28.1 & 28.3 & 28.5 & 28.5 & 28.5 & - & 28.0 \\
        PAN-13:14$+$24:31 & - & - & 27.4 & 27.7 & 27.9 & 28.3 & 28.3 & - & 27.9 \\
        PAN-13:43$+$55:49 & - & - & 27.7 & 28.0 & 28.2 & 28.5 & 28.5 & - & 28.0 \\
        PAN-14:25$+$56:30 & - & - & 28.3 & 28.6 & 28.8 & 29.0 & 29.0 & - & 28.5 \\
        PAN-14:49$+$10:17 & - & -& 27.9 & 28.1 & 28.3 & 28.5 & 28.5 & - & 28.1 \\
        PAN-15:34$+$23:25 & - & -& 27.7 & 28.0 & 28.2 & 28.5 & 28.6 & - & 28.2 \\
        PAN-17:07$+$58:52 & - & - & 27.3 & 27.6 & 27.8 & 28.2 & 28.2 & - & 27.8 \\
        PAN-22:16$+$00:24 & - & - & 29.6 & 29.5 & 29.6 & 29.3 & 29.2 & 28.9 & 28.9 \\
        PAN-22:16$+$00:25 & - & - & 29.6 & 29.5 & 29.7 & 29.2 & 29.1 & 28.8 & 28.8 \\
        PAN-22:46$-$05:18 & - & -& 28.6 & 28.6 & 28.2 & 28.5 & 28.8 & - & 28.6 \\
        PAN-22:46$-$05:31 & - & -& 29.3 & 29.4 & 29.5 & 29.5 & 29.5 & 28.8 & 29.3 \\
        PAN-23:45$-$42:35 & - & - & 28.8 & 28.7 & 28.5 & 28.6 & 28.9 & 28.2 & 28.4 \\
        BEA-01:04$+$02:15 & - & 28.2 & 28.5 & 28.9 & 29.1 & 29.0 & 29.0 & 28.4 & 28.5 \\
        BEA-00:14$-$30.33 & - & 28.2 & 28.5 & 28.8 & 28.9 & 29.0 & 29.1 & 28.3 & 28.5 \\
        BEA-02:17$-$05:03 & - & 29.2 & 29.3 & 29.3 & 29.2 & 29.5 & 29.3 & - & 29.1 \\
        BEA-04:46$-$26:36 & - & 27.7 & 27.7 & 28.3 & 28.4 & 28.7 & 28.6 & 27.9 & 28.2 \\
        BEA-05:01$-$43:37 & - & - & 27.7 & 28.0 & 28.2 & 28.5 & 28.5 & - & 28.1 \\
        BEA-08:42$+$03:24 & - & - & 27.9 & 28.1 & 28.3 & 28.3 & 28.4 & - & 28.0 \\
        BEA-09:59$+$02:00 & - & - & 28.0 & 28.2 & 28.9 & 28.6 & 28.9 & 28.3 & 28.2 \\
        BEA-12:27$+$21:56 & - & 28.5 & 28.7 & 28.9 & 29.1 & 29.1 & 29.0 & 28.4 & 28.6 \\
        BEA-12:29$+$27:02 & - & 26.6 & 26.3 & 26.6 & 26.8 & 27.4 & 27.4 & 27.0 & 27.2\\
        BEA-13:29$+$47:07 & - & 26.6 & 26.9 & 27.2 & 27.3 & 27.9 & 28.0 & 27.2 & 27. 7\\
        BEA-20:58$-$42:47 & - & - & 27.3 & 27.5 & 27.7 & 28.0 & 28.1 & - & 27.6 \\
        BEA-23:03$-$62:50 & - & 27.3 & 27.3 & 28.5 & 28.6 & 29.0 & 29.0 & 27.9 & 28.2 \\
        BEA-23:16$-$59:09 & - & - & 27.4 & 27.7 & 27.8 & 28.3 & 28.3 & - & 27.9 \\
        BEA-23:25$-$12:03 & - & 27.9 &  28.2 & 28.5 & 28.6 & 28.8 & 28.9 & 28.2 & 28.5 \\
        BEA-23:25$-$12:15 & - & - & 26.7 & 27.0 & 27.2 & 27.7 & 27.8 & - & 27.5 \\
        POP-07:40$+$31:04 & - & - & 26.9 & 27.2 & 27.4 & 27.4 & 27.5 & - & 27.1 \\
        POP-09:12$+$52:54 & - & - & 28.1 & 28.3 & 28.5 & 28.4 & 28.5 & - & 28.0 \\    
        POP-12:44$-$00:33 & - & - & 26.0 & 26.3 & 27.6 & 27.0 & 27.2 & - & 26.8 \\
        POP-13:50$+$47:53 & - & - & 28.1 & 28.4 & 28.5 & 28.5 & 28.5 & - & 28.1 \\
        \hline
    \end{tabularx}
    \label{table: depths2}
\end{table*}

\begin{table*}
    \centering
    \caption{Continued from Table \ref{table: depths}-\ref{table: depths2}, but also listing medium-band F210M and F300M 5$\sigma$ limiting magnitudes, as these filters were included for the fields below.}
    \begin{tabularx}{\textwidth}{l|XXXXXXXXXXXXXX}        \hline
        Field & F090W & F115W & F150W & F200W & F210M & F277W & F300M & F356W & F410M & F444W\\
        \hline
        MACS 0257-2325 & 28.1 & 28.2 & 28.1 & 28.3 & 28.1 & 28.4 & 28.3 & 28.4 & 28.1 & 28.2\\
        MACS 0257-2209$^{\dag}$ & 28.1 & 28.2 & 28.1 & 28.2 & 28.0 & 28.3 & 28.2 & 28.4 & 28.1 & 28.2\\
        MACS 0329 & 28.0 & 28.1 & 28.0 & 28.2 & 28.0 & 28.3 & 28.2 & 28.3 & 28.0 & 28.1 \\
        MACS 0308 & 27.9 & 28.0 & 28.0 & 28.1 & 27.9 & 28.1 & 28.1 & 28.2 & 27.9 & 28.0 \\
        MACS 1931 & 27.9 & 28.0 & 27.9 & 28.1 & 27.9 & 28.1 & 28.1 & 28.2 & 28.0 & 28.1 \\
        CLJ 0152 & 28.1 & 28.2 & 28.1 & 28.3 & 28.1 & 28.4 & 28.3 & 28.5 & 28.1 & 28.3 \\
        SMACS 2031$^{\dag}$ & 28.0 & 28.1 & 28.0 & 28.2 & 28.0 & 28.3 & 28.2 & 28.3 & 28.1 & 28.2 \\
        RXJ 0232 & 28.1 & 28.2 & 28.1 & 28.2 & 28.0 & 28.4 & 28.3 & 28.4 & 28.1 & 28.1 \\
        MACS 0451+0006$^{\dag}$ & 28.0 & 28.1 & 28.0 & 28.2 & 27.9 & 28.2 & 28.2 & 28.3 & 28.0 & 28.1 \\
        MS 0451-0305$^{\dag}$ & 28.0 & 28.1 & 28.0 & 28.2 & 28.0 & 28.3 & 28.2 & 28.3 & 28.0 & 28.1 \\
        Abell 383 & 28.0 & 28.1 & 28.0 & 28.2 & 28.0 & 28.3 & 28.2 & 28.3 & 28.0 & 28.2 \\
        Abell 3192 & 28.1 & 28.2 & 28.1 & 28.2 & 28.0 & 28.4 & 28.3 & 28.4 & 28.1 & 28.2 \\
        PLCKG 171 & 28.0 & 28.0 & 28.0 & 28.1 & 27.9 & 28.2 & 28.1 & 28.2 & 28.0 & 28.1 \\
        MACS 0429 & 28.0 & 28.1 & 28.1 & 28.2 & 28.0 & 28.3 & 28.2 & 28.4 & 28.1 & 28.2 \\
        SPT 2325$^{\dag}$ & 28.1 & 28.1 & 28.0 & 28.2 & 28.0 & 28.3 & 28.2 & 28.4 & 28.1 & 28.2 \\
        PLCKG 004 & 27.8 & 27.9 & 27.8 & 27.9 & 27.8 & 28.0 & 27.9 & 28.0 & 27.8 & 27.9 \\
        RXJ 2211 & 28.0 & 28.1 & 28.0 & 28.1 & 27.9 & 28.2 & 28.1 & 28.3 & 28.0 & 28.1 \\
        MACS 2214$^{\dag}$ & 28.0 & 28.1 & 28.0 & 28.2 & 28.0 & 28.3 & 28.2 & 28.3 & 28.1 & 28.2 \\
        SMACS 2131$^{\dag}$ & 28.0 & 28.1 & 28.0 & 28.2 & 28.0 & 28.3 & 28.2 & 28.3 & 28.0 & 28.0 \\
        MS 2137 & 28.0 & 28.0 & 27.9 & 28.1 & 27.9 & 28.2 & 28.1 & 28.2 & 27.9 & 28.0 \\
        MACS 2129 & 28.0 & 28.0 & 28.0 & 28.1 & 27.9 & 28.3 & 28.2 & 28.3 & 28.0 & 28.1 \\
        Abell 2537 & 28.0 & 28.1 & 28.0 & 28.2 & 28.0 & 28.2 & 28.1 & 28.3 & 28.0 & 28.1 \\
        MACS 0717 & 28.0 & 28.1 & 28.0 & 28.1 & 27.9 & 28.2 & 28.1 & 28.2 & 28.0 & 28.1 \\
        AbellS 295 & 28.1 & 28.2 & 28.1 & 28.3 & 28.1 & 28.4 & 28.3 & 28.5 & 28.1 & 28.3 \\
        MACS 0035 & 28.0 & 28.1 & 28.1 & 28.2 & 28.0 & 28.4 & 28.2 & 28.4 & 28.1 & 28.2 \\
        Abell 2813 & 28.0 & 28.1 & 28.0 & 28.2 & 27.9 & 28.2 & 28.1 & 28.3 & 28.0 & 28.1 \\
        Abell 209 & 28.0 & 28.1 & 28.0 & 28.2 & 28.0 & 28.2 & 28.2 & 28.3 & 28.0 & 28.1 \\
        Abell 68$^{\dag}$ & 28.0 & 28.1 & 28.0 & 28.2 & 28.0 & 28.3 & 28.2 & 28.3 & 28.0 & 28.2 \\
        RXJ 0600 & 28.4 & 28.5 & 28.6 & 28.5 & 28.0 & 28.5 & 28.1 & 28.6 & 28.2 & 28.4 \\
        RXJ 0032 & 28.1 & 28.2 & 28.1 & 28.2 & 28.0 & 28.3 & 28.2 & 28.4 & 28.1 & 28.2 \\
        MACS 0159 & 28.1 & 28.2 & 28.1 & 28.2 & 28.0 & 28.4 & 28.3 & 28.4 & 28.1 & 28.2 \\        
        TNJ 1338 & - & - & 27.8 & - & 27.2 & - & 27.8 & - & - & - \\
        MACS 0417Par & 29.1 & 29.4 & 29.5 & - & 29.1 & 29.6 & 28.9 & - & 29.1 & 29.2 \\
        MACS 1423Par & 29.1 & 29.3 & 29.4 & - & 29.0 & 29.6 & 28.9 & - & 29.0 & 29.2 \\
        MACS 1149Par & 29.1 & 29.3 & 29.6 & - & 29.0 & 29.6 & 28.9 & - & 29.0 & 29.2 \\
        MACS 0416Par & 29.1 & 29.3 & 29.5 & 29.1 & 29.0 & 29.5 & 28.9 & 29.3 & 29.0 & 29.2 \\
        \hline
    \end{tabularx}
    \label{table: depths3}
\end{table*}

% Don't change these lines
\bsp	% typesetting comment
\label{lastpage}

\end{document}